\documentstyle[12pt,epsfig]{article}
\def\llgm{\left\lgroup\matrix}
\def\rrgm{\right\rgroup}

\def\l{\ell}

\def\H{{\cal H}}

\def\be{\begin{equation}}
\def\ee{\end{equation}}
\def\ba{\begin{eqnarray}}
\def\ea{\end{eqnarray}}
\def\nnb{\nonumber}
\def\ed{\end{document}}
\def\rt{\right}
\def\dgr{\dagger}
\def\lt{\left}
\newcommand{\wti}{\widetilde}

\begin{document}

\renewcommand{\thefootnote}{\fnsymbol{footnote}}
\begin{titlepage}
\begin{flushright}
\begin{tabular}{l}
TUHEP-TH-00119\\
hep-ph/0006250
\end{tabular}
\end{flushright}
\vskip0.5cm
\begin{center}
{\LARGE\bf
$B_s \to \ell^+ \ell^-$ in a model II 2HDM and MSSM}
\vspace*{0.5cm}

       {\bf Chao-Shang HUANG}$^a$\footnote{E-mail : csh@itp.ac.cn},
	{\bf LIAO Wei }$^a$\footnote{E-mail: liaow@itp.ac.cn},
{\bf Qi-Shu YAN}$^b$\footnote{E-mail : qsyan@mail.tsinghua.edu.cn},
  and 
       {\bf Shou-Hua ZHU}$^{a,c}$\footnote{
	E-mail: huald@particle.uni-karlsruhe.de}
        \vspace{0.5cm}

        $^a$Institute of Theoretical Physics,
         Academia Sinica, 100080 Beijing, China \\
        $^b$ Physics Department of Tsinghua University, 100080 Beijing, China \\
    $^c$ Institut f\"ur Theoretische Physik, Universit\"at Karlsruhe,
        D-76128 Karlsruhe, Germany\\
\bigskip
  {\bf Abstract\\[10pt]}
\end{center}

In this paper  we analyze the process $B_s \to \ell^+ \ell^-$ in a
model II 2HDM and MSSM. All the leading terms of Wilson coefficients relevant
to the process are given in the large tan$\beta$ limit. It is shown that the 
decay width for $B_s \to \ell^+ \ell^-$ depends on all parameters except
$m_{A^0}$ in the 2HDM. The branching ratio of $B_s \to \mu^+ \mu^-$ can reach 
its experimental bound in some large tan$\beta$ regions of the parameter
space in MSSM because the amplitude increases like tan$^3\beta$ in the regions.
For l=$\tau$, the branching ratio can even reach $10^{-4}$ in the regions.
Therefore, the experimental measurements of leptonic decays of $B_s$ could
put a constraint on the 
contributions of neutral Higgs bosons and consequently the parameter space 
in MSSM.
\noindent
\vfill
\bigskip
\centerline{{\sc Pacs} numbers: 13.20.He, 13.25.Hw}
\end{titlepage}
\renewcommand{\thefootnote}{\arabic{footnote}}
\setcounter{footnote}{0}

\section{\bf Introduction}
$B_s\rightarrow l^+l^-$, as one of flavor changing neutral current processes, is 
sensitive to  structure of the standard model (SM) and 
new physics beyond SM, and is expected to shed light on the existence of
new physics before the possible new particles are produced at colliders.
Theoretically,
the process is clean  because only the nonperturbative quantity involved
is of the decay constant of $B_s$ and it is relatively easy to be
calculated
by so far well-known nonperturbative methods such as QCD sum rules, lattice gauge
theory, Bethe-Salpeter approach, etc. Therefore, it provides a good window to
probe new physics. Experimentally, the 95\% confidence level upper bound
on the 
$B_s \to \mu^{+} \mu^{-}$ branching fraction has been given \cite{cdf}:
\begin{equation}
	B_r (B_s \rightarrow \mu^{+}\mu^-) < 2.6 \times 10^{-6} 
\end{equation}
The planned experiments at B-factories are likely to measure branching
fractions as low as $10^{-8}${\cite{ali1}}. 

Compared to the rare decay $B\rightarrow X_s \gamma$,  $B_s\rightarrow
l^+l^-$ (as well as
$B\rightarrow X_s l^+l^-$) is of more advantage for the study of the Higgs
sector in the large 
tan$\beta$ case in a model II two Higgs doublet model (2HDM) or supersymmetric models (SUSY) since the 
contributions to $B\rightarrow X_s \gamma$ coming from Higgs sector are indeed independent of tan$\beta$ when  
tan$\beta$ is larger than a few ( say, 4 ).

The branching ratio for  $B_s\rightarrow l^+l^-$  has been calculated in SM and beyond SM
in a number of papers~\cite{hnv,sk,cg}. In a recent paper~\cite{ln} the process in a model II
2HDM with large tan$\beta$ is reanalyzed. It is correctly pointed out that the contributions
of the box diagram to this decay at the leading order of tan$\beta$ are
missed and a minus for the 
contribution of $A^0$ penguin diagram involving $H^{\pm}$ and $W^{\pm}$ in the loop is also
missed in the earlier literature~\cite{hnv,sk,dhh}. However, there are some points in the paper which need to 
be clarified. First, the argument that the trilinear $H^{\pm}H^{\mp}H$( H=$h^0,H^0$) couplings 
should not be considered as tan$\beta$ enhanced is not correct, as we shall argue below. Second, 
although the contribution of box diagram is the same order as those
of penguin diagrams in the large tan$\beta$ limit it is numerically smaller than those of
penguin diagrams and consequently the claim that the box diagram gives the dominant contribution in the
't Hooft-Feynman gauge is not true. In the paper we shall give a detailed argument (as we know,
there is no such argument presented in the literature ) to show why their
claim on $H^{\pm}H^{\mp}H$ couplings is not correct (see section 3). These arguments are important to
clarify where are the disagreements in the literature and make one have a correct conclusion on 
the decay in 2HDM.

There are more box diagrams in SUSY than that in 2HDM.  
The contributions of box diagrams in the analysis in supersymmetric models are missed in the previous 
papers~\cite{hy,hly,cg}. The contributions are also omitted in the refs.\cite{mas,goto}
since they neglect the mass of a lepton in calculating Wilson coefficients. However, for l=$\mu, 
\tau$ the contributions in the large tan$\beta$ case are important and consequently should not be neglected. In the paper we calculate the contributions to 
Wilson coefficients $C_9, C_{Q_i}$ from the box diagrams and carry 
out a complete analysis in a model II 2HDM and SUSY with large tan$\beta$.

The paper is organized as follows. Section 2 is devoted to the effective Hamiltonian 
responsible for $b\rightarrow s l^+l^-$. We calculate the contributions to Wilson coefficients from the box
diagrams and give all the leading terms of Wilson coefficients in a model
II 2HDM and SUSY with large tan$\beta$ in section 3. In section 4 we present the numerical
results. In section 5 conclusions are drawn. Finally the
contributions to Wilson coefficients $C_9, C_{Q_i}$ from individual diagrams in a 2HDM and 
MSSM are given in the appendix.

\section{ Effective Hamiltonian for $b\rightarrow s \ell^+ \ell^-$}
The effective Hamiltonian describing the flavor changing processes
$b\rightarrow s \ell^+ \ell^-$ can be defined as
\ba
\label{halm}
H_{eff}&=&-\frac{G_F}{\sqrt{2}} \lambda_t (\sum_{i=1}^{10}C_i(\mu)O_i(\mu)
+\sum_{i=1}^{10}C_{Q_i}(\mu)Q_i(\mu)).
\ea
where $\lambda_t=V_{tb} V_{ts}^*$, $O_i$s $(i=1,\cdots,10)$ are the 
same as those given 
in the ref.\cite{wise,buras}\footnote{We follow the convention in
ref.\cite{wise} for the indices of operators as well as Wilson
coefficients. In the convention of ref.\cite{buras} $O_8$ ($O_9$) is
changed into $O_9$ ($O_{10}$).}, and
$Q_i$s come from exchanging neutral Higgs bosons and have been
given in refs. ~\cite{dhh,hy}.

The QCD corrections to coefficients $C_i$ and $C_{Q_i}$ can be incorporated
in the standard way by using the renormalization group equations. $Q_i(i=1,\cdots,10)$ does 
not mix with $O_8,O_9$ so that the evolution of $C_8$ and $C_9$
remains unchanged and are given in ref.\cite{wise}
\begin{eqnarray}
C_8(m_b)&=&C_8(m_W)+\frac{4\pi}{\alpha_s(m_W)}[-\frac{4}{33}(1-\eta^{-11/23})
+\frac{8}{87}(1-\eta^{-29/23})]C_2(m_W),\nnb\\
&&=C_8(m_W)+1.937 C_2(m_W)\\
C_9(m_b)&=&C_9(m_W).
\end{eqnarray}

It is obvious that operators $O_i(i=1,\cdots,10)$ and $Q_i(i=3,\cdots,10)$
do not mix into $Q_1$ and $Q_2$ and also there is no mixing between $Q_1$
and $Q_2$. Therefore, the evolution of $C_{Q_1},C_{Q_2}$ is controlled by the 
anomalous dimensions of $Q_1,Q_2$ respectively.
\begin{eqnarray}
C_{Q_i}(m_b)&=&\eta^{-\gamma_Q/\beta_0}C_{Q_i}(m_W),~~i=1,2,\nnb\\
&=&1.24 C_{Q_i}(m_W)
\end{eqnarray}
where $\gamma_Q=-4$ \cite{h} is the one loop anomalous dimension of $\bar{s}_Lb_R$,
$\eta = \frac{\alpha_s(m_b)}{\alpha_s(M_W)}\approx
1.72$, and $\beta_0 = 11 - (2/3)n_f = 23/3$.  

For the decay $B_s\rightarrow l^{+}l^-$, the matrix element of ${\cal H}_{eff}$ is to be 
taken between vacuum and $|B^0_s\rangle$ state. Because
\be
\label{four}
\langle 0 |~ {\bar s} ~\sigma^{\mu \nu} ~P_R ~b~ |B^0_s\rangle = 0
\ee
and  the $O_8$ term in eq.(\ref {halm})
gives zero on contraction with the lepton bilinear due to
 $ p^\mu_B = p^\mu_+ + p^\mu_-$, 
 only the operators $O_{9}$ and $Q_i$
\begin{eqnarray}
O_9=\bar{s}_L\gamma^{\mu}b_L \bar{l}\gamma_{\mu}\gamma_5 l,\\
O_{Q_{1}}= \bar{s}_Lb_R \bar{l} l, \\
O_{Q_{2}}= \bar{s}_Lb_R \bar{l}\gamma_5 l 
\end{eqnarray}
are involved 
and the important thing we need to do is to calculate the Wilson coefficients of
the operators at $\mu=m_W$. $C_9 (m_W)$ has been calculated in SM \cite{il}, in a 2HDM
\cite{wise} and in SUSY models~\cite{mas,goto} respectively. $C_9(m_W)$ in the 2HDM 
is the same as that in SM for the large $tan\beta$ scenario. The box
diagram contributions
to $C_9(m_W)$ (as well as $C_8(m_W)$)
in SUSY which are proportional to tan$^2\beta$ are missed in the previous 
calculations~\cite{goto,hy,hly,cg}. $C_{Q_i}$s have also been 
calculated in 2HDM~\cite{dhh,sk,ln} and in SUSY~\cite{hy,hly}. However,
some leading terms in the large tan$\beta$ limit are missed in the previous papers.
We shall calculate $C_9$ and $C_{Q_i}$ at $\mu=m_W$ in 
the next section in order to give a complete and correct result.

By using the equations of motion for quark fields, we have
\be
\label{three}
\langle 0~| {\bar s} ~\gamma_5~ b| B^0_s \rangle 
= i f_{B_s}\frac{ m^2_{B_s}}{m_b+m_s} 
\ee
 where $f_{B_s}$ is the decay constant of $B_s$ defined by
\be
\label{two}
\langle 0~|~ {\bar s} ~\gamma^\mu~ \gamma_5 ~b~| B^0_s \rangle 
= -i f_{B_s} p^\mu \\
\ee
\par Thus 
the effective Hamiltonian ($\ref{halm}$) results in the following decay 
amplitude for $B_s\rightarrow l^+l^-$
\begin{equation}
	A = \frac{G_F \alpha_{EM}}{ 2 \sqrt{2} \pi} 
	m_{B_s} f_{B_s}
	 \lambda_t
	\left[ \ C_{Q_1}  \bar{u}  v
	+ \left(  C_{Q_2}
	+2{\hat m_l} C_9 \right) 
	\bar{u} \gamma_5 v \right],
\end{equation}
where ${\hat m_l}=m_l/m_{B_s}$ and $m_{B_s}\approx m_b+m_s$ have been used.
Then it is straightforward to obtain the branching ratio for $B_s\rightarrow l^+l^-$
\begin{eqnarray}
    Br~ (B_{s} \to \ell^+\ell^-) 
	&=& \frac{G_F^2 \alpha_{EM}^2}{64 \pi^3}
	m_{B_{s}}^3 \tau_{B_{s}} f^2_{B_{s}}
	 |\lambda_t|^2
	\sqrt{1 - 4 {\hat m_{\ell}^2}}
	\nonumber \\
	&& \times
	\left[ \left(1 - 4 {\hat m_{\ell}^2}\right)
	 |C_{Q_1}|^2
	+ | C_{Q_2} 
	+2 {\hat m_{\ell}} C_9 |^2 \right],	
\end{eqnarray}
where $\tau_{B_{s}}$ is the $B_{s}$ lifetime.

\section{\bf Wilson coefficients}
\subsection{In 2HDM}
Consider two complex hypercharge $Y=1$, $SU(2)_w$ doublet scalar fields, $\phi_1$ and
$\phi_2$. The Higgs potential which spontaneously breaks $SU(2)\times U(1)$
down to $U(1)_{EM}$ and conserves CP symmetry can be written in the following form 
~\cite{Georgi}:
\begin{eqnarray}
\label{poten}
V(\phi_1,\phi_2) &=&
\lambda_1 (\Phi_1^{\dag} \Phi_1- v_1^2)^2
 +\lambda_2  (\Phi_2^{\dag} \Phi_2 -v_2^2)^2 \nonumber\\
&&+ \lambda_3 [ (\Phi_1^{\dag} \Phi_1 -v_1^2)+
(\Phi_2^{\dag} \Phi_2- v_2^2) ]^2
\nonumber\\
&&+ \lambda_4 [ (\Phi_1^{\dag} \Phi_1) (\Phi_2^{\dag} \Phi_2)-
(\phi_1^{\dag} \Phi_2) (\Phi_2^{\dag} \Phi_1) ]
\nonumber \\
&&
+ \lambda_5
 [ \mbox{Re}(\Phi_1^{\dag} \Phi_2)- v_1 v_2 ]^2
\nonumber \\
&&
+ \lambda_6
 [ \mbox{Im}(\Phi_1^{\dag} \Phi_2)]^2
\label{eq2}
\end{eqnarray}
Hermiticity requires that all parameters are real. If $\lambda_i\geq 0$ 
the potential is semi-positive and the minimum of the potential is at
\begin{eqnarray}
<\Phi_1>=\left( \begin{array}{c}
0 \\
v_1
\end{array}
\right), \ \ \ \
<\Phi_2>=\left( \begin{array}{c}
0 \\
v_2 
\end{array}
\right),
\end{eqnarray}
thus breaking $SU(2)\times U(1)$ down to $U(1)_{EM}$.
 From the potential the mass eigenstates are easily found as follows.
The charged Higgs states are
\begin{eqnarray}
	G^{\pm} &=& \phi_1^{\pm} \cos\beta + \phi_2^{\pm} \sin\beta \nonumber \\
	H^{\pm} &=& -\phi_1^{\pm} \sin\beta + \phi_2^{\pm} \cos\beta,
\end{eqnarray}
where the mixing angle $\beta$ is defined by tan$\beta$=$v_2$/$v_1$.  The CP--odd states are
\begin{eqnarray}
	G^0 &=& \sqrt{2}[Im \phi_1^{0} \cos\beta + Im \phi_2^{0} \sin\beta] \nonumber \\
	A^0 &=& \sqrt{2}[-Im \phi_1^{0} \sin\beta + Im \phi_2^{0} \cos\beta],
\end{eqnarray}
The would--be Goldstone bosons $G^{\pm}$ and $G^0$ are eaten 
by the $W$ and $Z$ bosons. The physical Higgs boson masses are
\begin{eqnarray}
m_{H^{\pm}}^2=\lambda_4 v^2,\\
m_{A^0}^2= \lambda_6 v^2,
\end{eqnarray}
where $v^2=v^2_1+v^2_2$ is fixed by the $W$ boson mass, 
$M^2_W = \frac{1}{2}g^2 v^2$.
Diagonalizing the mass matrix of CP-even Higgs 
\be
v^2 \llgm{4 cos^2\beta (\lambda_1+\lambda_3)+sin^2\beta \lambda_5&(4 \lambda_3+\lambda_5)sin2\beta/2\cr
(4 \lambda_3+\lambda_5)sin2\beta/2 &4 sin^2\beta (\lambda_2+\lambda_3)+cos^2\beta \lambda_5}\rrgm,
\ee
results in the CP--even eigenstates 

\begin{eqnarray}
	H^0 &=& \sqrt{2}[(Re \phi_1^0 - v_1) \cos\alpha + (Re \phi_2^0 - v_2) \sin\alpha], \nonumber \\
	h^0 &=& \sqrt{2}[(-Re \phi_1^0 - v_1) \sin\alpha + (Re \phi_2^0 - v_2) \cos\alpha]
\end{eqnarray}
with masses
\ba
\label{m1}
m^2_{h^0}+m^2_{H^0}&=&v^2 [4(cos^2\beta \lambda_1+sin^2\beta \lambda_2)+ \lambda_+]\;,\\
\label{m2}
m^2_{H^0}-m^2_{h^0}&=&v^2 [ \lambda_- cos2\beta - 4 \lambda_2 sin^2\beta  + 4 \lambda_1
 cos^2\beta ] / cos2\alpha
\ea
and mixing angle
\begin{equation}
\label{m3}
tan 2\alpha=\frac{\lambda_+ sin2\beta}{\lambda_- cos2\beta
			-4 \lambda_2 sin^2\beta+4 \lambda_1 cos^2\beta }\;
\end{equation}
where $\lambda_{\pm}=4 \lambda_3\pm \lambda_5$.
In the potential ($\ref{poten}$) there are 8 parameters : $\lambda_i$, i=1,...,6, $v_1$
and $v_2$. Because, as said above, $v^2$ is fixed by $m_W$, there are seven
independent parameters in a general CP invariant 2HDM. Six of them can be
expressed in terms of mixing angles $\alpha$ and $\beta$ and Higgs masses
$m_{H^{\pm}}, m_{A^0}, m_{H^0}, m_{h^0}$. The seventh needs to be fixed by
measuring one of the quartic coupling in ($\ref{poten}$). For simplicity we shall assume
$\lambda_1=\lambda_2$ hereafter so that we have six independent parameters
in the model. Taking $\lambda_1=\lambda_2$, eqs.($\ref{m1},\ref{m2},\ref{m3}$) reduce to
\ba
m^2_{H^0}+m^2_{h^0}&=& v^2 (4\lambda_1 + \lambda_{+})\;,\\
m^2_{H^0}-m^2_{h^0}&=& 
v^2 \sqrt{(4\lambda_1+\lambda_-)^2 cos^2 2\beta+ \lambda_+^2 sin^2 2\beta}\;\\
tan2\alpha&=&\frac{\lambda_+}{\lambda_{-}+4 \lambda_1}tan2\beta.
\ea
The three equations show explicitly that the angle $\alpha$ as well as masses
$m_{H^0}, m_{h^0}$ can be traded for $\lambda_i$, i=1,3,5 (or, equivalently, 1,+,-)
no matter how large tan$\beta$ is. That is, $\alpha$, as one of the set of six independent 
parameters which contains both $\alpha$ and $\beta$ as well as others,
 can take any value independent of tan$\beta$, as it should be. Therefore, the statement in
ref. ~\cite{ln} that the angle $\alpha$ depends on $\beta$ is not correct. When tan$\beta$
approaches to infinity, if (4$\lambda_1+\lambda_-)$ and consequently $m_{H^0}^2-m_{h^0}^2$
is of order cot$\beta$, say, (4$\lambda_1+\lambda_-)$= c tan2$\beta$ with c a constant of
order one., then sin2$\alpha$ is of order one. If (4$\lambda_1+\lambda_-)$ is of order one,
then in the large tan$\beta$ limit it follows that $sin\alpha \sim cot\beta$ or 1-$cot^2\beta$
/2 so that sin$2\alpha$ goes always as cot$\beta$ which cancels the
tan$\beta$ 
enhancement. However, the conclusion is valid only at tree level. Once the
radiative corrections are 
included it would change, which is similar to the situation that happens in the Higgs
sector of MSSM, i.e., the radiative corrections violate the tree level mass relations and
one treats Higgs boson masses as free parameters to be determined by experiments. Therefore,
even in this case we should still treat $\alpha$ as well as $\beta$ and Higgs
masses as free parameters in the general 2HDM defined above so that the 
tan$^2\beta$ enhancement due to the trilinear $H^{\pm}H^{\mp}H$( H=$h^0,H^0$) couplings should 
be considered, as we did in ref.~\cite{dhh}.

As usual, in the model II 2HDM  the
Higgs--fermion Yukawa couplings are given by 
\begin{equation}
	L_{Yuk} = - Y_D \bar{Q} \Phi_1 D 
	- Y_U \bar{Q} \Phi_2^c U - Y_l \bar{L} \Phi_1 l + {\rm h.c.}
\end{equation}
where $\Phi^c = i \tau_2 \Phi^*$. So down--type
quarks and charged leptons (up--type quarks) acquire masses by their 
couplings to $\Phi_1$ ($\Phi_2$).

 Feynman rules in the above general 2HDM have been given. Vertices with one
or two gauge bosons and vertices involving two fermions and one boson are given in
ref.~\cite{ghkd}. The three Higgs boson vertices can be found in ref.~\cite{sk}.
 The vertices involving one Goldstone boson and two Higgs bosons
have also been given in ref.~\cite{ghkd}. We use these Feynman rules in calculations
of Wilson coefficients.

As pointed out in section II, for large $tan\beta$, $C_9(m_W)$ in the 2HDM is 
the same as that in SM. The leading contributions to $C_{Q_i}$ in the large
tan$\beta$ limit come from the diagrams in Fig. 1. In our previous paper~\cite{dhh}
we paid attention to the contributions of neutral Higgs bosons and missed the
contribution from the box diagram involving one charged Higgs and one W boson which
is of order tan$^2\beta$ in the large tan$\beta$ limit~\cite{ln}. We carry
out a calculation for the diagram and confirm the result in
ref.~\cite{ln}. In this
paper we include the contribution and correct a sign for $A^0$ penguin.
In order to separate contributions from individual diagrams, we write $C_{Q_i}$ as
\ba
C_{Q_i}&=&C_{Q_i}^{S}+C_{Q_i}^{B}+C_{Q_i}^{P},
\ea
where
$C_{Q_i}^{S}$, $C_{Q_i}^{B}$, and $C_{Q_i}^{P}$ denote the contributions from self-energy 
type diagrams, box diagrams, and Higgs penguin diagrams, respectively.
In appendix A we present all contributions proportional to tan$^2\beta$ in Feynman-t'Hooft
gauge. Adding all tan$^2\beta$ contributions together, we have
\ba
\label{cq}
C_{Q_1}(m_W)&=&f_{ac} y_t \bigg[ \frac{lny_t}{1-y_t}- \frac{sin^2(2\alpha)}{4}
\frac{(m_{h^0}^2-m_{H^0}^2)^2}{m_{h^0}^2 m_{H^0}^2} f_1(y_t) \bigg], \\
C_{Q_2}(m_W)&=&-f_{ac} y_t \frac{lny_t}{1-y_t}.\\
C_{Q_3}(m_W)&=&\frac{m_be^2}{m_{\tau}g^2}(C_{Q_1}(m_W)+C_{Q_2}(m_W)),\\
C_{Q_4}(m_W)&=&\frac{m_be^2}{m_{\tau}g^2}(C_{Q_1}(m_W)-C_{Q_2}(m_W)),\\
C_{Q_i}(m_W)&=&0, ~~~~i=5,\cdots, 10
\ea
where
$$
f_{ac}=\frac{m_b m_l tan^2\beta}{4 sin^2\theta_W m_W^2},
x_t=\frac{m^2_t}{m^2_W},~~y_t=\frac{m^2_t}{m^2_{H^{\pm}}},
f_1(y)=\frac{1-y+ylny}{(y-1)^2}.
$$
The difference between eq.($\ref{cq}$) and the result in ref.~\cite{ln} is that the 
first term in the brackets in eq.($\ref{cq}$) is incorrectly omitted in
ref.~\cite{ln}.
It is worth to note that  in the above equations $m_b$=$m_b (m_W)$.
\subsection{In SUSY}
In the minimal supersymmetric standard model (MSSM) or supergravity
model (SUGRA) the Higgs sector is the same as that in a model II 2HDM by
imposing the following constraints on the parameters~\cite{ghkd}:
\ba
\lambda_2=\lambda_1 \\
\lambda_3=\frac{1}{8}(g^2+g^{'2})-\lambda_1 \label{con1} \\ 
\lambda_4=2\lambda_1-\frac{1}{2}g^{'2} \label{con2} \\
\lambda_5=\lambda_6=-2(\lambda_1+2\lambda_3). \label{con3}
\ea
And all Feynman rules can be found in ref.~\cite{ghkd}. In addition to Fig. 1,
the diagrams in Fig. 2 also give the leading contributions. 
Besides box diagrams, five different sets of contributions to the decay
$b\rightarrow s l^+l^-$ are present in SUSY. They can be
classified
according to the virtual particles exchanged in the loop: {\it a)} the SM
contribution with exchange of $W^-$ and up-quarks; {\it b)} the charged
Higgs boson contribution with $H^-$ and up-quarks; {\it c)} the chargino
contribution with $\chi^-$ and up-squarks ($\wti u$);
{\it d)} the gluino contribution with $\tilde g$ and down-squarks
($\wti d$); and finally {\it e)} the neutralino contribution
with $\chi^0$ and down-squarks.
 As pointed out
in refs.~\cite{mas,bo,hly}, contributions from neutrilino-down type squark (e) and
gluino-down type squark (d) loop diagrams are negligible compared to those 
from chargino-up type squark diagrams because the flavor mixings between the third and the other
two generations are small in minimal supergravity and constrained MSSM. Therefore, 
in addition to the SM (a) and charged Higgs (b) contributions, we only include
the contributions from chargino-up type squark (c) loop diagrams in the paper.

In some regions of the parameter space the dominant contribution to $C_{Q_i}$ is proportional to 
tan$^3\beta$ and comes from the self-energy type diagrams, as pointed out in ref.~\cite{hy,hly,bk}.
 In quite a large region of the parameter space the dominant contribution is proportional to 
tan$^2\beta$. Box diagrams can contribute terms with tan$^4\beta$ to $C_{Q_i}$ which
are greatly suppressed by $(m_s/M_W)^2$, therefore the largest
contributions to $C_{Q_i}$ from 
SUSY box diagrams remain proportional to tan$^2\beta$. Among the diagrams
in Fig.2 only the box diagram 
with charginos in the loop can give the tan$^2\beta$ enhancement to
$C_9(m_W)$. The contributions of the self 
energy type and penguin diagrams to $C_{Q_i}$ have been calculated by us
in refs.~\cite{hy,hly}. We 
calculate the contributions of the box diagrams and summarize
all contributions in the appendix B. Adding all contributions given in the appendix B, 
one has
\ba
&& C_{Q_1}(M_W) = f_{ac} \{ -\frac{sin^2(2\alpha)}{2} \frac{(m_{h^0}^2-m_{H^0}^2)^2}
{2 m_{h^0}^2 m_{H^0}^2} y_t f_1(y_t)- (x_t -1) f_{C^0}(x_{H^-},1,x_t) \ \nonumber \\
&& - f_{C^0}(x_{H^-},1,0) \}- f_{ac} \frac{M_W}{m_b \lambda_t} \sum_{i,j=1}^2 \sum_{k,k'=1}^6 \H_{jbk'}
\Gamma^\dgr_{isk} \bigg\{ \delta_{ij} \delta_{kk'} r_{hH} \sqrt{x_{\chi_i^-}}
f_{B^0}\lt(x_{\chi_i^-}, x_{\tilde u_k}\rt) \ \nonumber \\
&&- \frac{\sqrt{2}}{tan\beta} [\delta_{kk'} G^{hH}_{ijk} 
f_{C^0}(x_{\chi_i^-},x_{\chi_j^-},x_{{\tilde u}_k}) + \delta_{ij} F^{hH}_{kk'}
\sqrt{x_{\chi_i^-}} f_{C^0}(x_{\chi_i^-},x_{\tilde u_k},x_{\tilde u_{k'}})] \ \nonumber \\
&& + \frac{1}{tan\beta} \delta_{kk'} \lt( \sqrt{x_{\chi_i^-} x_{\chi_j^-}} Q_{ij}+x_{\tilde u_k} 
Q_{ij}^\dgr \rt)f_{D^0}(x_{\chi_i^-},x_{\chi_j^-},x_{\tilde u_k},x_{\tilde \nu_l})\bigg\} \, \label{c1} \\
&&=- tan^3\beta \frac{m_b m_\ell}{4 sin^2\theta_w M_W^2 \lambda_t} \sum_{i=1}^2 \sum_{k=1}^6 U_{i2} 
T^{km}_{UL} K_{m b} \{-\sqrt{2} V_{i1} (T_{UL} K)^*_{ks}+ V_{i2} \frac{ (T_{UR} 
{\tilde m_u} K)^*_{ks}}{M_W sin\beta} \}\nnb \\ &&r_{hH} \sqrt{x_{\chi_i^-}} f_{B^0}\lt(x_{\chi_i^-}, 
x_{\tilde u_k}\rt) +O(tan^2\beta),\label{c1a}\\
&& C_{Q_2}(M_W) = f_{ac} \bigg[(x_t -1) f_{C^0}(x_{H^-},1,x_t) + f_{C^0}(x_{H^-},1,0)
\bigg] \ \nonumber\\
&& + f_{ac} \frac{M_W}{m_b \lambda_t} \sum_{i,j=1}^2 \sum_{k,k'=1}^6 \H_{jbk'} \Gamma^\dgr_{isk} \bigg\{ \delta_{ij} \delta_{kk'}
r_{A} \sqrt{x_{\chi_i^-}} f_{B^0}\lt(x_{\chi_i^-}, x_{\tilde u_k}\rt) \ \nonumber \\
&&+ \frac{\sqrt{2}}{tan\beta} [\delta_{kk'} G^A_{ijk} 
f_{C^0}(x_{\chi_i^-},x_{\chi_j^-},x_{{\tilde u}_k}) + \delta_{ij} F^A_{kk'}
\sqrt{x_{\chi_i^-}} f_{C^0}(x_{\chi_i^-},x_{\tilde u_k},x_{\tilde u_{k'}})] \ \nonumber \\
&& - \frac{1}{tan\beta} \delta_{kk'} \lt( \sqrt{x_{\chi_i^-} x_{\chi_j^-}} Q_{ij}-x_{\tilde u_k} 
Q_{ij}^\dgr \rt)f_{D^0}(x_{\chi_i^-},x_{\chi_j^-},x_{\tilde u_k},x_{\tilde \nu_l})\bigg\} \,\label{c2} \\
&&=tan^3\beta \frac{m_b m_\ell}{4 sin^2\theta_w M_W^2 \lambda_t} \sum_{i=1}^2 
\sum_{k=1}^6 U_{i2} T^{km}_{UL} K_{m b} \{-\sqrt{2} V_{i1} (T_{UL} K)^*_{ks}+ V_{i2} \frac{(T_{UR} 
{\tilde m_u} K)^*_{ks}}{M_W sin\beta}\}\nnb \\ && r_{A} \sqrt{x_{\chi_i^-}} f_{B^0}\lt(x_{\chi_i^-}, 
x_{\tilde u_k}\rt) +O(tan^2\beta),\label{c2a}\\
&& C_{9} (M_W) = -\frac{1}{4 sin^2\theta_W} \bigg \{ x_t F_9(x_{t})- G(x_t,0)+ G(0,0)
+ \frac{y_t}{2} ctg^2\beta \bigg( F_2(y_t)+ F_4(y_t)\bigg) \ \nonumber \\
&& -\sum_{i,j=1}^2 \sum_{k,k'=1}^6 \Gamma_{jbk'} \Gamma_{isk}^\dgr \bigg[
\frac{1}{4} \bigg ( \delta_{ij} \sum_{m=1}^{3} T^{km}_{UL} T^{\dgr mk'}_{UL} G_0(x_{\tilde u_k \chi_j^-},x_{\tilde u_{k'} \chi_j^-}) \ \nnb \\
&& + 2 \delta_{kk'} \sqrt{x_{\chi_j^- \tilde u_{k'}} x_{\chi_i^- \tilde u_k}} U_{j1}^* U_{i1} F_0(x_{\chi_j^- \tilde u_k},x_{\chi_i^- \tilde u_k}) 
- \delta_{kk'} V_{j1} V_{i1}^* \bigg [ log(x_{\tilde u_k}) + G_0(x_{\chi_j^- \tilde u_k},x_{\chi_i^- \tilde u_k}) \bigg ] \bigg ) \ \nnb \\
&&- \frac{1}{2} \delta_{kk'} \bigg ( \frac{1}{x_{\chi_j^-}} V_{j1}V_{i1}^* G^{\prime}(x_{{\tilde u_k} \chi_j^-},x_{\tilde \nu_{\tau} \chi_j^-},x_{\chi_i^- \chi_j^-})
\nnb\\ &&
+ \frac{m_{\ell}^2}{M_W^2} tan^2\beta \sqrt{x_{\chi_j^-} x_{\chi_i^-}} U_{j2}^* U_{i2}
 f_{D^0}(x_{\chi_i^-},x_{\chi_j^-},x_{\tilde u_k},x_{\tilde \nu_l}) \bigg ) \bigg] \bigg\},\label{c3}
\ea
where $U$ and $V$ are matrices which diagonalize the mass matrix of charginos,
$T_{Ui}$ (i=L, R) is the matrix which diagonalizes the mass matrix of the scalar up-type quarks and K is the CKM matrix. For
the definitions of various symbols in the above equations, see the Apendex B. In eqs. (\ref{c1a})
and (\ref{c2a}) the tan$^3\beta$ term has been explicitly written.\\
Let us give some remarks: 

(a) The first term in eq. (\ref{c1}) which is propotional to sin$^22\alpha$ arises from the trilinear 
$H^{\pm}H^{\mp}H$ couplings. At tree level, due to the more constraints (\ref{con1}-\ref{con3}) than that in the 2HDM
defined in subsection 3.1, there are only two free parameters in the Higgs sector of MSSM which we may choose as
tan$\beta$ and one of masses of Higgs bosons, e. g., $m_{A^0}$. In the large tan$\beta$ limit one has
\ba
m_{h^0}^2&\approx & m_Z^2, ~~~~~~~~~m_{H^0}\approx m_{A^0}^2, ~~~~~~~~~~~~m_{H^{\pm}}^2=m_{A^0}^2+m_W^2, \nnb\\
sin 2\alpha &\sim & cot\beta. \label{sin}
\ea
So the first term would have no tan$\beta$ enhencement. However, including the radiative corrections, the above relations are
, in general, changed and the mixing angle $\alpha$ is determined by
\ba
tan2\alpha=\frac{sin 2\beta (m_{A^0}^2+m_Z^2) - 2 R_{12}}{cos 2\beta ( m_{A^0}^2-m_Z^2) +R_{22}-R_{11}}, ~~~~ - \frac{\pi}
{2} < \alpha \leq 0, \label{alpha}
\ea
where $R_{ij}$ are the radiative corrections to the mass matrix of the neutral Higgs bosons in the $\{ H_1^0, H_2^0 \}$ basis and
have been given in refs.\cite{oyy,cqw}.  As shown in ref.\cite{hcsz}, $R_{12}$ can reach more than ten percents of $R_{22}$ in
the case of $\mu\sim A_t\sim A_b$ and consequently sin 2$\alpha$ can be the order of one.  Of course, there are some cases
in which $R_{12}$ as well as $R_{11}$ is very small and of a few thousandth of $R_{22}$ and consequently the tree level relations
(\ref{sin}) are almost not changed. Therefore, we keep, in general, $\alpha$ and Higgs boson masses as free parameters because
several parameters in MSSM enter the Higgs sector through the raditive corrections. 

(b) The first terms in eqs. (\ref{c1a}, \ref{c2a} ) which arise from the self-energy type diagram will provide the tan$^3\beta$ enhencement,
as pointed out in~\cite{hy,hly}, 
if the mass splittings of stops are large (say, ${\geq}$100Gev).  The condition is necessary because if all the squark 
masses are  degenerate ($m_{\tilde{t}_1}=m_{\tilde{t}_2}=\tilde{m}$), the large contributions arising from
the chargino-squark loop exactly cancel due to the GIM mechanism\cite{bsg4}. We remark that the 
chirality structure of the $Q_i$(i=1, 2) operators allows a large tan$\beta$ enhancement
for the Wilson coefficients $C_{Q_i}$(i=1, 2), as happened
for the magnetic moment operator O$_7$,
and there is no such a large tan$\beta$ enhancement for the Wilson
coefficients $C_i$(i=8, 9)
due to the different chirality structure of the $O_i$(i=8, 9)
operators. 

(c)The last term in eq. (\ref{c3}) comes from the chargino-chargino box diagrams and are proportional to tan$^2\beta$. It was missed
in the literature. For l=$\mu, \tau$ and large tan$\beta$, it is numerically the same size as the other contributions so that it should not be omitted.

\section{\bf Numerical results}

Below we assume no CP violating phases from 2HDM and SUSY. As said in section II, there
are 6 free parameters in the 2HDM which are $tan\beta$, $\alpha$, $m_{h^0}$,
$m_{H^0}$, $m_{A^0}$, $m_{H^\pm}$. In MSSM, in addition to the above 6 parameters, 7 extra
free parameters, $m_{\tilde c_L}$, $m_{\tilde t_L}$, $m_{\tilde t_R}$, $A_t$, $M_2$,
$\mu$, and $m_{\tilde \nu_{\tau}}$ are needed in order to calculate the Wilson coefficients.
In Table. 1 we list all SM inputs for our numerical analysis.

Numerical results are given in Figs. 3-9. Figs. 3-5 are devoted to the decay
$B_s \rightarrow \mu^+ \mu^-$ in 2HDM. In the numerical calculations in 2HDM the constraint on $m_{H^{\pm}}$ from
$b\rightarrow s \gamma$\cite{ahmed} has been imposed. We present in Fig. 3 the branching ratio
(Br) for $B_s \rightarrow \mu^+ \mu^-$  as the function of $m_{H^0}$, the mass
of the heavier CP even neutral Higgs boson, for fixed values of the other parameters.
The figure shows that the Br increases when $m_{H^0}$ increases except for the mixing
angle $\alpha$=0. The reason is that in the large tan$\beta$ case the trilinear $H^{\pm}H^{\mp}H$ (H=$H^0,h^0$)
couplings are proportional to $m_{H^0}^2$ and/or $m_{h^0}^2$ and the couplings vanish when $\alpha$
is equal to zero. One can see from Fig. 4 that the Br, as the function of $\alpha$,
behaves like sin$^{2}2\alpha$ when $m_{H^0}$ is large enough ( say, 500 GeV), which
implies the contributions from the trlinear couplings of NHBs dominate
(see eqs. (2.13), (3.30)
and (3.31)). Fig. 5 shows the tan$\beta$ dependence of Br and one can see
from it that the
contributions coming from $C_{Q_i}^{~~'}$s can dominate when tan$\beta$ is large enough (say, 
larger than 80). In most of the large tan$\beta$ region in the parameter space Br is
of order $10^{-8}$, an order of magnitude larger than that in SM.

\begin{table}
\begin{center}
\begin{tabular}{|c|c|c|c|c|c|c|}
\hline
 $m_b$ & $m_c$ & $m_{\mu}$ & $m_{\tau}$ & $M_{B_s}$ & $f_{B_s}$ & $G_F$ \\
\hline
$4.8$ GeV & $1.4$ GeV & $0.11$ GeV & $1.78$ GeV & $5.37$ GeV & 0.22 GeV & $1.17\times 10^{-5}$GeV$^{-2}$ \\
\hline
 $\alpha^{-1}$ & $|V^\ast_{ts} V_{tb}|$ &  $|V_{cb}|$ & $Br(b\rightarrow c e \nu_e)$ & $\tau_{B_s}$ & $sin^2\theta_w$ & $\alpha_s(M_z)$ \\
\hline
 $ 129$ & $0.0385$ & $0.036$ & $0.114$ & $1.54\times 10^{-12}$ s &
$0.232$ & $0.12$ \\
 \hline
  \end{tabular}
 \end{center}
\caption{\it Values of the standard model parameters used in our
numerical analysis.}
\end{table}

\begin{table}
        \begin{center}
        \begin{tabular}{|c|c|c|c|c|c|c|}
        \hline
 $\alpha$ & $m_{h^0}$ & $m_{H^0}$ & $m_{A^0}$ & $m_{H^-}$ & $m_{\chi_1^-}$ & $m_{\tilde \nu_{\tau}}$ \\
\hline
 $0.08$ & $110$ GeV & $160$ GeV & $160$ GeV & $180$ GeV & $90$ Gev & $150$ GeV \\
\hline
        \end{tabular}
        \end{center}
\caption{\it }
\label{tab:para}
\end{table}

The numerical results in MSSM are presented in Figs. 6-9. 
We present the correlation between $C_7$ and $C_{Q_1}$ for $l=\mu$ 
and $l=\tau$ respectively in Fig. 6 and 7 where the absolute values of $C_7$ are taken from the data
of $B \to X_s \gamma$\cite{ahmed} with the $2\sigma$ errors imposed. 
We set
$m_{\tilde t_1}$, $m_{\tilde t_2}$, $m_{\tilde c_L}$, $m_{\chi_2^-}$, and
$tan\beta$ as random free parameters. They vary in the range 180-300 GeV,
250-450 GeV, 200-400 Gev, 160-360 Gev and 25-50 respectively.
Other parameters are fixed as given in Table 2.
We get about 3000  permitted points among 25000 points.
The contributions to Wilson
coefficients due to superparticles in a loop ( SUSY contributions) come
mainly through the ${\tilde u_k}^* \bar \chi_i d$ vertex,
which is determined by the mixing between Higgsinos and Winos and the mixing between stops. The vertex appears 
in Feynman diagrams which describe the processes $b\rightarrow s\gamma$ and $b\rightarrow s l^+l^-$ so that
there exists a correlation between $C_7$ and $C_{Q_i}$. In some large tan$\beta$ regions of the parameter space in MSSM,
SUSY contributions interfere destructively with the SM contributions and SUSY contributions can be so large that they can overwhelm those from the SM and the 
Higgs sector so that the sign of $C_7$ is changed compared to that in SM, as shown in Fig. 6 and 7. In the regions $C_{Q_i}$s
are proportional to tan$^3\beta$ and consequently can compete with $C_9$ for l=$\mu$ and be much larger than $C_9$ for l=$\tau$. We also calculate the 
correlation between $C_{Q_1}$ and $C_{Q_2}$ in the regions and it follows
that $C_{Q_1}\approx -C_{Q_2}$.

The contribution to $C_9$ which is proportional to $tan^2\beta$ coming 
from chargino-chargino box diagrams is numerically the same order as other
contributions from chargino-chargino box or chargino-up type squark
penguin diagrams. As a whole SUSY contributions to $C_9$ give about $10\%$
corrections to the SM value. Taking $C_{Q_1}$ in the allowed range
in Figs. 6 and 7 and $C_{Q_2}\approx -C_{Q_1}$, we draw the branching ratio
of $B_s\rightarrow l^+l^-$ as function of $C_{Q_1}$ in Figs. 8 and 9, given $C_9$ being the SM value with
$10\%$ variations. Figs. 8 is for l=$\mu$ and Fig. 9 
for l=$\tau$. From Figs. 8,9, we can see that Br($B_s\rightarrow l^+ l^-$)
is more sensitive to $C_{Q_i}^{~~'}$s (which represent the contributions from NHBs ) than to $C_9$ in the large tan$\beta$ (larger than about 30) regions 
because in the regions $|C_{Q_i}^{~~'}|$s are much larger than $|\frac{m_l}{m_{B_s}}C_9|$. The numerical results tell us that it is possible to saturate
the experimental bound (1.1) for $B_s\rightarrow \mu^+\mu^-$ in some
regions of the parameter
space in MSSM. In other words, the experimental bound could impose
a constraint on the parameter space of SUSY which we shall analyze elsewhere. Because $C_{Q_i}^{~~'}$s are proportional to the
lepton mass, Br for l=$\tau$ can reach order of $10^{-4}$ in the regions in which Br($B_s\rightarrow \mu^+\mu^-$) saturates
the experimental bound. 

\section{\bf Conclusions}
In summary we have analyzed the decays $B_{s} \to \ell^+ \ell^-$
in the model II 2HDM and SUSY with large $\tan\beta$.  Although these decays
have been studied in these models before and reanalyzed recently, it seems that no
complete analysis exists so far. We have calculated all leading terms in the large
tan$\beta$ limit. We found that in addition to the Higgs boson-W boson box diagram,
the chargino-chargino box diagram gives also a contribution proportional to tan$^2\beta$,
the former to $C_{Q_i}$ (i=1,2) and the latter to $C_i$ (i=8,9). The contributions from NHBs
always increase the branching ratios in the large $tan\beta$ case so that the branching ratios in the 2HDM and in SUSY are larger than those in SM. 
We have numerically computed the branching ratios for l=$\mu$ and $\tau$. In the
2HDM the branching ratio for l=$\mu$ is about $10^{-8}$, an order of
magnitude larger than that in SM,  if 
tan$\beta$ = 50 or so and the other parameters are in the reasonable
range. We have shown the dependence
of the branching ratio with respect to the mixing angle $\alpha$ and
neutral Higgs boson masses. 
The branching ratio increases when the splitting of the masses of the two
CP even neutral Higgs bosons increases except for
the case of the mixing angle $\alpha$=0.
 In MSSM the branching ratio for l=$\mu$ can saturate the experimental
bound
in some regions of the parameter space 
where $C_{Q_i}$s (i=1,2) behave as tan$^3\beta$. In the other regions where $C_{Q_i}$s (i=1,2) 
behave as tan$^2\beta$ the branching ratio is about the order $10^{-8}$.
The branching ratio for l=$\tau$ reaches $10^{-4}$ in the regions
of the parameter space in which Br($B_s\rightarrow l^+l^-$) saturates the experimental bound. In the near future when 
very high statistics
can be reached~\cite{ali1,fe} the  measurements of the decays
$B_s\rightarrow
l^+l^-$ (l=$\mu, \tau$)  can provide a large potential to find or
exclude the large tan$\beta$ parts of the parameter space in 2HDM  and/or SUSY.

\vskip1cm

\section*{\bf Acknowledgements}\nonumber
One of the authors (C.-S. Huang) would like to thank Gad Eilam for discussions and Department
of Physics, Technion-Israel Institute of Technology  where a part of work was done for warm hospitality. 
This work was supported in part by the National Nature Science Foundation of China and supproted in part by  the Alexander von Humboldt Foundation.

\section*{\bf Appendix}\nonumber
\subsection*{A Wilson coefficients in a model II 2HDM}
Wilson coefficients are extracted from the transition amplitudes
by integrating out heavy particles. In our convention, the effective Hamiltonian
is related with the amplitude by
\be
i M(b \to s l^+ l^-) = -i < l^+ l^- s| H_{eff} |b>
\ee
By computating the self-energy type, Higgs-penguin and box diagrams, $C^i_{Q_1}$ and 
$C^i_{Q_2}$ with the superscript denoting the type of a diagram are extracted out, 
as given below
\begin{eqnarray}
C_{Q_1}^{S}&=& f_{ac} r_{hH} (x_{H^-}-1) x_t f_2(x_t,y_t)\;,\\
C_{Q_2}^{S}&=&- f_{ac} r_A (x_{H^-}-1) x_t f_2(x_t,y_t) \;,\\
C_{Q_1}^{P}&=& -f_{ac}\frac{sin^22\alpha}{2} \frac{(m_{h^0}^2-m_{H^0}^2)^2}{2 m_{h^0}^2 m_{H^0}^2} y_t f_1(y_t)\nonumber\\
&-&f_{ac} \lt[-1+(x_{H^-}-1) r_{hH}\rt]x_t f_2(x_t,y_t) \;,\\
C_{Q_2}^{P}&=& f_{ac} \lt[-1+(x_{H^-}-1) r_A\rt] x_t f_2(x_t,y_t) \;,\\
C_{Q_1}^{B}&=&-f_{ac} B_+ \! \lt( x_{H^+}, x_t \rt)\;,\\
C_{Q_2}^{B}&=&f_{ac} B_+ \! \lt( x_{H^+}, x_t \rt)\;,
\end{eqnarray}
where
\be
f_{ac}=\frac{1}{4 sin^2\theta_w} \frac{m_b m_{\l} }{M_W^2} tan^2\beta\;,
r_{hH}=M_W^2 \lt(\frac{\sin^2 \alpha}{M_{h^0}^2} + \frac{\cos^2 \alpha}{M_{H^0}^2}\rt)\;,
\ee
\be
r_A=\frac{M_W^2}{m_{A^0}^2}\;,x_{H^-}=\frac{m_{H^-}^2}{M_W^2}\;, 
x_{t}=\frac{m_t^2}{M_W^2}\;, y_t=\frac{m^2_t}{m^2_{H^-}}, 
\ee
\ba
f_1(y)=\frac{1-y+ylny}{(y-1)^2}, \\
f_2(x,y)=\frac{xlny}{(z-x)(x-1)}+\frac{lnz}{(z-1)(x-1)},
~~\textrm{with} ~~z=x/y,\\
B_+(x,y)=\frac{y}{x-y}(\frac{lny}{y-1}-\frac{lnx}{x-1}),
\ea

\subsection*{B Wilson coefficients in MSSM}
\subsubsection*{B.1 Feynman rules and conventions}
In this subsection we present our convention. 
In order to avoid the trouble in dealing with charge conjugate operation, we
choose $\chi^-$ as the particle.
The interactions of $d{\tilde u}\chi$, $H\chi\chi$, $H{\tilde u}{\tilde u}$, and
$Hdd$ can be expressed as:
\ba
{\cal L}_{d{\tilde u}\chi}&=&\frac{g_2}{\sqrt{2}} {\tilde u_k^*} {\bar \chi_i^-}[\Gamma_{ijk}\;P_L+\H_{ijk}\;P_R] d_j +{\it h.c.}\;,\\
{\cal L}_{H\chi\chi}&=&\frac{g_2}{\sqrt{2}}{\bar\chi_i^-} [g^-_{ijh}P_L+g^+_{ijh} P_R] \chi_j^- H^0_h+{\it h.c.}\;,\\
{\cal L}_{H{\tilde u}{\tilde u}}&=&\frac{g_2}{\sqrt{2}} \; f_{ijh}\; {\tilde u_i}^* {\tilde u_j} H^0_h+{\it h.c.}\;,\\
{\cal L}_{Hdd}&=&\frac{g_2}{\sqrt{2}} \; g_h \; {\bar d} d H^0_h \;,
\ea
where
\ba
\Gamma_{ijk}&=&-\sqrt{2} V_{i1}^* (T_{UL} K)_{kj}+ V_{i2}^* \frac{ (T_{UR} {\tilde m_u} K)_{kj}}{M_W sin\beta}\;,\\
\H_{ijk}&=&U_{i2} \frac{(T_{UL} K {\tilde m_d})_{kj}}{M_W cos\beta}\;,\\
g^+_{ijh}&=&\sqrt{2} \lt \{ (Q_{ij}sin\alpha-S_{ij} cos\alpha),-(Q_{ij} cos\alpha+S_{ij}sin\alpha),-i(Q_{ij}sin\beta+S_{ij}cos\beta) \rt\}_h\;,\\
g^{-}_{ijh}&=&\sqrt{2} \lt\{ (Q_{ij}^\dgr sin\alpha-S_{ij}^\dgr cos\alpha),-(Q_{ij}^\dgr cos\alpha +S_{ij}^\dgr sin\alpha),i(Q_{ij}^\dgr sin\beta+S_{ij}^\dgr cos\beta)\rt \}_h\;,\\
g_{h}&=&\frac{m_d}{\sqrt{2} M_W}\{\frac{sin\alpha}{cos\beta},-\frac{cos\alpha}{cos\beta},i\; tan\beta \gamma_5\}_h\;,
\ea
\be
f_{ijh}=T_{il}H^h_{ll'}T_{l'j}^{\dgr}\;,
Q_{ij}=\frac{1}{\sqrt{2}} U_{i2} V_{j1}\;,
S_{ij}=\frac{1}{\sqrt{2}} U_{i1} V_{j2}\;.
\ee
In (7.58) and (7.59) $K$ is the CKM matrix, and ${\tilde m_u}$ and ${\tilde m_d}$ are defined as
\be
{\tilde m_u}=diag\{m_u,m_c,m_t\}\;,{\tilde m_d}=diag\{m_d,m_s,m_b\}\;.
\ee
$H^h$ can be expressed as
\be
H^{h0}= -\llgm{\sqrt{2}\lt[ \frac{cos\alpha}{sin\beta}\frac{{\tilde m_u}^2}{M_W}-\frac{sin(\alpha+\beta)}{cos^2\theta_w} D_{uL} M_W\rt ]& \frac{{\tilde m_u}}{\sqrt{2} M_W}\lt(\frac{sin\alpha}{sin\beta}\mu 1 +\frac{cos\alpha}{sin\beta}A^\dgr\rt) \cr
\frac{1}{\sqrt{2} M_W}\lt(\frac{sin\alpha}{sin\beta}\mu^* 1+\frac{cos\alpha}{sin\beta}A\rt) {\tilde m_u} &\sqrt{2}\lt[ \frac{cos\alpha}{sin\beta}\frac{{\tilde m_u}^2}{M_W}-\frac{sin(\alpha+\beta)}{cos^2\theta_w} D_{uR} M_W\rt ]}\rrgm,\\
\ee
\be
H^{H0}= -\llgm{\sqrt{2}\lt[ \frac{sin\alpha}{sin\beta}\frac{{\tilde m_u}^2}{M_W}+\frac{cos(\alpha+\beta)}{cos^2\theta_w} D_{uL} M_W\rt ]& \frac{{\tilde m_u}}{\sqrt{2} M_W}\lt(-\frac{cos\alpha}{sin\beta}\mu 1+\frac{sin\alpha}{sin\beta}A^\dgr\rt) \cr
 \frac{1}{\sqrt{2} M_W}\lt(-\frac{cos\alpha}{sin\beta}\mu^* 1 +\frac{sin\alpha}{sin\beta}A \rt) {\tilde m_u} &\sqrt{2}\lt[ \frac{sin\alpha}{sin\beta}\frac{{\tilde m_u}^2}{M_W}+\frac{cos(\alpha+\beta)}{cos^2\theta_w} D_{uR}M_W\rt ]}\rrgm,\\
\ee
\be
H^{A0}=-\llgm{ 0& \frac{- i {\tilde m_u}}{\sqrt{2} M_W}\lt(\mu 1+ctg\beta A^\dgr \rt) \cr
 \frac{i}{\sqrt{2} M_W}\lt(\mu^* 1 +A ctg\beta \rt) {\tilde m_u} &0}\rrgm\;.
\ee
where
\be
D_{uL}=(\frac{1}{2}-e_u sin^2\theta_w) \;, D_{uR}=e_u sin^2\theta_w\;.
\ee

The $6\times6$ mass matrix of u-type squark is given as
\be
{\cal M}_{\tilde u}=
   \llgm{\lt( m_{\tilde Q}^2 +{\tilde m_u}^2\rt) \;+cos2\beta D_{uL}m_z^2 & {\tilde m_u} \lt(-\mu ctg\beta 1 + A^\dgr \rt) \cr
\lt(-\mu^* ctg\beta 1+ A\rt) {\tilde m_u} &m_{\tilde U}^2 +{\tilde m_u}^2 \;+cos2\beta D_{uR} m_z^2 }\rrgm,
\ee
where each block is a $3\times3$ matrix. $A$ is defined by $Y^u=A h^u$, while $Y^u$
is the trilinear coupling matrix of up-type squarks, and $h^u$ is the Yukawa coupling
of up-type quarks. The $6\times6$ $T$ matrix is defined as
\ba
diag{m^2_{\tilde u_k}}=T {\cal M}_{\tilde u} T^\dgr\;,{\tilde u_k}=T_{UL}^{kl} K_{li} {\tilde u_L^i} +T_{UR}^{ki'} {\tilde u_R^{i'}}.
\ea

The convention of chargino masses is given as
\ba
{\cal L}_{\chi\chi}&=&-\phi^{-T} X \phi^+ + {\it h.c.}\;,
\ea
\be
X= \llgm{ M_2 & \sqrt 2 M_W sin\beta \cr \sqrt 2 M_W cos\beta &\mu}\rrgm,
\ee
\be
\phi^{-T}=\lt\{-i\lambda^-,\phi_{H_1}^-\rt\}\;,\phi^{+T}=\lt\{-i\lambda^+,\phi_{H_2}^+\rt\}\;,
\ee
\be
\phi^-=U^+\chi^-_0\;,\phi^+=V^+\chi^+_0\;,\chi^{-T}_i=\{\chi^-_{0i},{\bar\chi^+_{0i}}\}.
\ee

where
\be
U^* X V^{\dgr}=diag\{m_{\chi_1^\pm}, m_{\chi_2^\pm} \}
\ee

With the above conventions, it is straightforward to extract Feynman
rules.

\subsubsection*{B.2 $C_{9}$, $C_{Q_1}$ and $C_{Q_2}$ in SUSY}
Wilson coefficients are extracted from the transition amplitudes
by integrating out heavy particles. $C_{9}$ is given as
\ba
C_{9}&=&C_{9,z}+C_{9,B}\\
C_{9,z}&=&C_{9,z}^W+C_{9,z}^{H^-}+C_{9,z}^{\chi^-}\\
C_{9,B}&=&C_{9,B}^W+C_{9,B}^{\chi^-}\\
C_{9,z}^W&=&-\frac{x_t}{4 sin^2\theta_W} F_9(x_{tW})\\
C_{9,z}^{H^-}&=&-\frac{x_t}{4 sin^2\theta_W} \frac{x_{th}}{2} ctg^2\beta [F_3(x_{th})+F_4(x_{th})]\\
C_{9,z}^{\chi^-}&=&\frac{1}{4 \lambda_t sin^2\theta_W} \frac{1}{4} \sum_{i,j=1}^2 \sum_{k,k'=1}^6 \Gamma_{jbk'} \Gamma_{isk}^\dgr\lt \{\delta_{ij} \sum_{m=1}^{3} T^{km}_{UL} T^{\dgr mk'}_{UL} G_0(x_{\tilde u_k \chi_j},x_{\tilde u_{k'} \chi_j}) \rt. \nnb\\
&&\lt.+\delta_{kk'} \lt [ 2 \sqrt{x_{\chi_j^- \tilde u_{k'}} x_{\chi_i^- \tilde u_k}} U_{j1}^* U_{i1} F_0(x_{\chi_j^- \tilde u_k},x_{\chi_i^- \tilde u_k})  \rt.\rt. \nnb\\
&&\lt. \lt.+V_{j1} V_{i1}^* \lt ( -log(x_{\tilde u_k}) - G_0(x_{\chi_j^- \tilde u_k},x_{\chi_i^- \tilde u_k}) \rt ) \rt ] \rt \}\\
C_{9,B}^W&=&\frac{1}{4 sin^2\theta_W}[G(x_{tW},0)-G(0,0)]\\
C_{9,B}^{\chi^-}&=&\frac{1}{4 \lambda_t sin^2\theta_W} \frac{1}{2} \sum_{i,j=1}^2\sum_{k=1}^6 \Gamma_{jbk} \Gamma^\dgr_{isk} \lt[ \frac{1}{x_{\chi_j^-}} V_{j1}V_{i1}^* G^{\prime}(x_{{\tilde u_k} \chi_j^-},x_{\tilde \nu_{\tau} \chi_j^-},x_{\chi_i^- \chi_j^-})\rt.\nnb\\
&&\lt.+ \frac{m_{\ell}^2}{M_W^2} tan^2\beta \sqrt{x_{\chi_j^-} x_{\chi_i^-}} U_{j2}^* U_{i2} f_{D^0}(x_{\chi_i^-},x_{\chi_j^-},x_{\tilde u_k},x_{\tilde \nu_{\tau}})\rt ]
\ea
By computating the self energy type, Higgs-penguin and box diagrams, $C^{ij}_{Q_1}$ and $C^{ij}
_{Q_2}$ with the first superscipt denoting the type of a diagram and the second superscript
a Higgs boson or a superparticle in the loop of the diagram
are extracted out, as given below
\ba
C_{Q_{1(2)}}^{S}&=&C_{Q_{1(2)}}^{SH}+C_{Q_{1(2)}}^{SS}\;,\\
C_{Q_{1(2)}}^{P}&=&C_{Q_{1(2)}}^{PH}+C_{Q_{1(2)}}^{PS}\;,\\
C_{Q_{1(2)}}^{B}&=&C_{Q_{1(2)}}^{BH}+C_{Q_{1(2)}}^{BS}\;,\\
C_{Q_1}^{SH}&=&-f_{ac} r_{hH} (x_{H^-}-1) x_t f_{C^0}(x_{H^-},1,x_t) \;,\\
C_{Q_2}^{SH}&=&f_{ac} r_A (x_{H^-}-1) x_t f_{C^0}(x_{H^-},1,x_t)\;,\\
C_{Q_1}^{PH}&=&-f_{ac}\frac{sin^22\alpha}{2} \frac{(m^2_{h^0}-m^2_{H^0})^2} {2 m^2_{H^0} m^2_{h^0}} y_t f_1(y_t)\nnb \\
&&+f_{ac} \lt[-1+(x_{H^-}-1) r_{hH}\rt] x_t f_{C^0}(x_{H^-},1,x_t)\;,\\
C_{Q_2}^{PH}&=&-f_{ac} \lt[-1+(x_{H^-}-1) r_A\rt] x_t f_{C^0}(x_{H^-},1,x_t) \;,\\
C_{Q_1}^{BH}&=&f_{ac} \lt[f_{C^0}(x_{H^-},1,x_t)-f_{C^0}(x_{H^-},1,0)\rt]\;,\\
C_{Q_2}^{BH}&=&-f_{ac} \lt[f_{C^0}(x_{H^-},1,x_t)-f_{C^0}(x_{H^-},1,0)\rt]\;,\\
C_{Q_1}^{SS}&=&-f_{ac} r_{hH} \frac{M_W}{m_b \lambda_t} \sum_{i=1}^2 \sum_{k=1}^6 \H_{ibk} \Gamma^{\dgr}_{isk} \sqrt{x_{\chi_i^-}}f_{B^0}\lt(x_{\chi_i^-}, x_{\tilde u_k}\rt)\;,\\
C_{Q_2}^{SS}&=&f_{ac} r_A \frac{M_W}{m_b \lambda_t} \sum_{i=1}^2 \sum_{k=1}^6 \H_{ibk} \Gamma^{\dgr}_{isk} \sqrt{x_{\chi_i^-}} f_{B^0}\lt(x_{\chi_i^-}, x_{\tilde u_k}\rt)\;,\\
C_{Q_1}^{PS}&=&f_{ac}  \frac{\sqrt{2} M_W}{m_b tan \beta \lambda_t} \sum_{i,j=1}^2 \sum_{k,k'=1}^6 \H_{jbk'} \Gamma^\dgr_{isk} \lt [ \delta_{kk'} G^{hH}_{ijk} f_{C^0}(x_{\chi_i^-},x_{\chi_j^-},x_{{\tilde u}_k})\rt.\nnb\\
&&\lt.+ \delta_{ij} F^{hH}_{kk'} \sqrt{x_{\chi_i^-}} f_{C^0}(x_{\chi_i^-},x_{\tilde u_k},x_{\tilde u_{k'}})\rt]\;,\\
C_{Q_2}^{PS}&=&f_{ac}  \frac{\sqrt{2} M_W}{m_b tan \beta \lambda_t} \sum_{i,j=1}^2 \sum_{k,k'=1}^6 \H_{jbk'} \Gamma^\dgr_{isk} \lt [ \delta_{kk'} G^{A}_{ijk} f_{C^0}(x_{\chi_i^-},x_{\chi_j^-},x_{{\tilde u}_k})\rt.\nnb\\
&&\lt.+ \delta_{ij} F^{A}_{kk'} \sqrt{x_{\chi_i^-}} f_{C^0}(x_{\chi_i^-},x_{\tilde u_k},x_{\tilde u_{k'}})\rt]\;,\\
C_{Q_1}^{BS}&=&-f_{ac} \frac{M_W}{m_b tan\beta \lambda_t} \sum_{i,j=1}^2 \sum_{k=1}^6 \H_{jbk} \Gamma^\dgr_{isk} \lt( \sqrt{x_{\chi_i^-} x_{\chi_j^-}} Q_{ij}+x_{\tilde u_k} Q_{ij}^\dgr \rt)f_{D^0}(x_{\chi_i^-},x_{\chi_j^-},x_{\tilde u_k},x_{\tilde \nu_l})\;,\\
C_{Q_2}^{BS}&=&-f_{ac} \frac{M_W}{m_b tan\beta \lambda_t} \sum_{i,j=1}^2 \sum_{k=1}^6 \H_{jbk} \Gamma^\dgr_{isk} \lt( \sqrt{x_{\chi_i^-} x_{\chi_j^-}} Q_{ij} - x_{\tilde u_k} Q_{ij}^\dgr \rt) f_{D^0}(x_{\chi_i^-},x_{\chi_j^-},x_{\tilde u_k},x_{\tilde \nu_l})\;,
\ea
where
\be
f_{ac}=\frac{1}{4 sin^2\theta_w} \frac{m_b m_{\l} }{M_W^2} tan^2\beta\;,
r_{hH}=M_W^2 \lt(\frac{\sin^2 \alpha}{M_{h^0}^2} + \frac{\cos^2 \alpha}{M_{H^0}^2}\rt)\;,
\ee
\be
r_A=\frac{M_W^2}{m_{A^0}^2}\;,x_{H^-}=\frac{m_{H^-}^2}{M_W^2}\;, x_{t}=\frac{m_t^2}{M_W^2}\;, x_{\chi_i^-}=\frac{m^2_{\chi_i^-}}{M_W^2}\;, x_{{\tilde u}_k}=\frac{m^2_{{\tilde u}_k}}{M_W^2}\;,
\ee
\ba
G^{hH}_{ijk}&=&\frac{M_W^2}{m_{h^0}^2} sin\alpha (\sqrt{x_{\chi_i^-} x_{\chi_j^-}} g^+_{ijh^0} + x_{\tilde u_k}g^-_{ijh^0}) - \frac{M_W^2}{m_{H^0}^2} cos\alpha (\sqrt{x_{\chi_i^-} x_{\chi_j^-}} g^+_{ijH^0} + x_{\tilde u_k} g^-_{ijH^0}) \;,
\ea
\be
G^{A}_{ijk}=i \frac{M_W^2}{m_{A^0}^2} (\sqrt{x_{\chi_i^-} x_{\chi_j^-}} g^+_{i j A^0} + x_{\tilde u_k}g^-_{i j A^0})\;,
\ee
\be
F^{hH}_{kk'}=\frac{M_W }{m_{h^0}^2} sin\alpha f_{kk' h^0} - \frac{M_W }{m_{H^0}^2} cos\alpha f_{kk' H^0}\;,
F^{A}_{kk'}=i \frac{M_W}{m_{A^0}^2} f_{kk'A^0}\;.
\ee
The related loop-integral functions are defined as
\ba
F_3(x)&=&\frac{1}{2 (x-1)^3} \lt[x^2-4 x+3+2 logx\rt]\;,\\
F_4(x)&=&\frac{1}{2 (x-1)^3} \lt[x^2-1-2 x \;logx\rt]\;,\\
F_9(x)&=&\frac{1}{2 (x-1)^2} \lt[x^2-7 x+6+(3 x+2)\; logx\rt]\;,\\
F_0(x,y)&=&\frac{1}{x-y} \lt [\frac{x}{x-1}logx-(x\rightarrow y)\rt]\;,\\
G_0(x,y)&=&\frac{1}{x-y} \lt [\frac{x^2}{x-1}logx-\frac{3}{2} x -(x\rightarrow y)\rt]\;,\\
G(x,y)&=&\frac{1}{x-y} \lt [\frac{x^2}{x-1}logx-\frac{1}{x-1} -(x\rightarrow y)\rt]\;,\\
G'(x,y,z)&=&\frac{1}{x-y} \lt [G_0(x,z)-G_0(y,z)\rt]\;,\\
f_{B^0}(x,y)&=&\frac{1}{x-y}\lt(x logx-ylogy\rt)\;,\\
f_{C^0}(x,y,z)&=&\frac{1}{y-z}\lt(f_{B^0}(x,y)-f_{B^0}(x,z)\rt)\;,\\
f_{D^0}(x,y,z,w)&=&\frac{1}{z-w}\lt(f_{C^0}(x,y,z)-f_{C^0}(x,y,w)\rt)\;.
\ea

In the above expressions for Wilson coefficients,  the possible
mixing among sneutrinos has been neglected.


Figure Captions\\

Fig. 1 Feynman diagrams which give the leading contributions to
$C_{Q_1}$ and $C_{Q_2}$ in the 2HDM with large tan$\beta$.

Fig. 2 Dominant Feynman diagrams in which the virtual superparticles are
exchanged in the loop in MSSM with large tan$\beta$. 

Fig. 3 The branching ratio of $B_s \to \mu^+ \mu^-$
as functions of $M_{H^0}$ in the 2HDM. Curves labelled by 1, 2, 3
corresponds to $\alpha = 0, \pi/8, \pi/4$
respectively. Other parameters are chosen to be
$M_{h^0} =120$ GeV, $M_{H^\pm} = 250$ GeV
and $tan\beta = 60$.

Fig. 4 The branching ratio as functions of $\alpha$ in the 2HDM.
Four curves labelled by 1, 2, 3, 4 corespond to $M_{H^0}
= 220, 320, 420, 520$ GeV respectively. Other parameters
are the same in Fig. 3

Fig. 5 The branching ratio as functions of $tan\beta$
with $\alpha = \pi/8$ in the 2HDM. Four curves are
classified as in Fig. 4. Other parameters are chosen as
in Fig. 3.

Fig. 6: The correlation between $C_7$ and $C_{Q_1}$ for l=$\mu$ in MSSM
 with $m_{\tilde t_1}$, $m_{\tilde t_2}$, $m_{\tilde c_L}$,
$m_{\chi_2^-}$ and $tan\beta$ as free parameters.

Fig. 7: The correlation between $C_7$ and $C_{Q_1}$ for l=$\tau$ in MSSM
 with $m_{\tilde t_1}$, $m_{\tilde t_2}$, $m_{\tilde c_L}$,
$m_{\chi_2^-}$ and $tan\beta$ as free parameters.

Fig. 8: $Br(B_s \rightarrow \mu^+ \mu^-)$  as a function
of $C_{Q1}$ in MSSM. $C_{Q1}\approx -C_{Q2}$ has been assumed.
Three lines correspond to $C_9 = -4.2, -4.6, -5.0$ respectively.

Fig. 9: $Br(B_s \rightarrow \tau^+ \tau^-)$  as a function
of $C_{Q1}$ in MSSM. $C_{Q1}\approx -C_{Q2}$ has been assumed.
Three lines correspond to $C_9 = -4.2, -4.6, -5.0$ respectively.

\begin{figure}
\begin{minipage}[t]{6.1 in}
     \epsfig{file=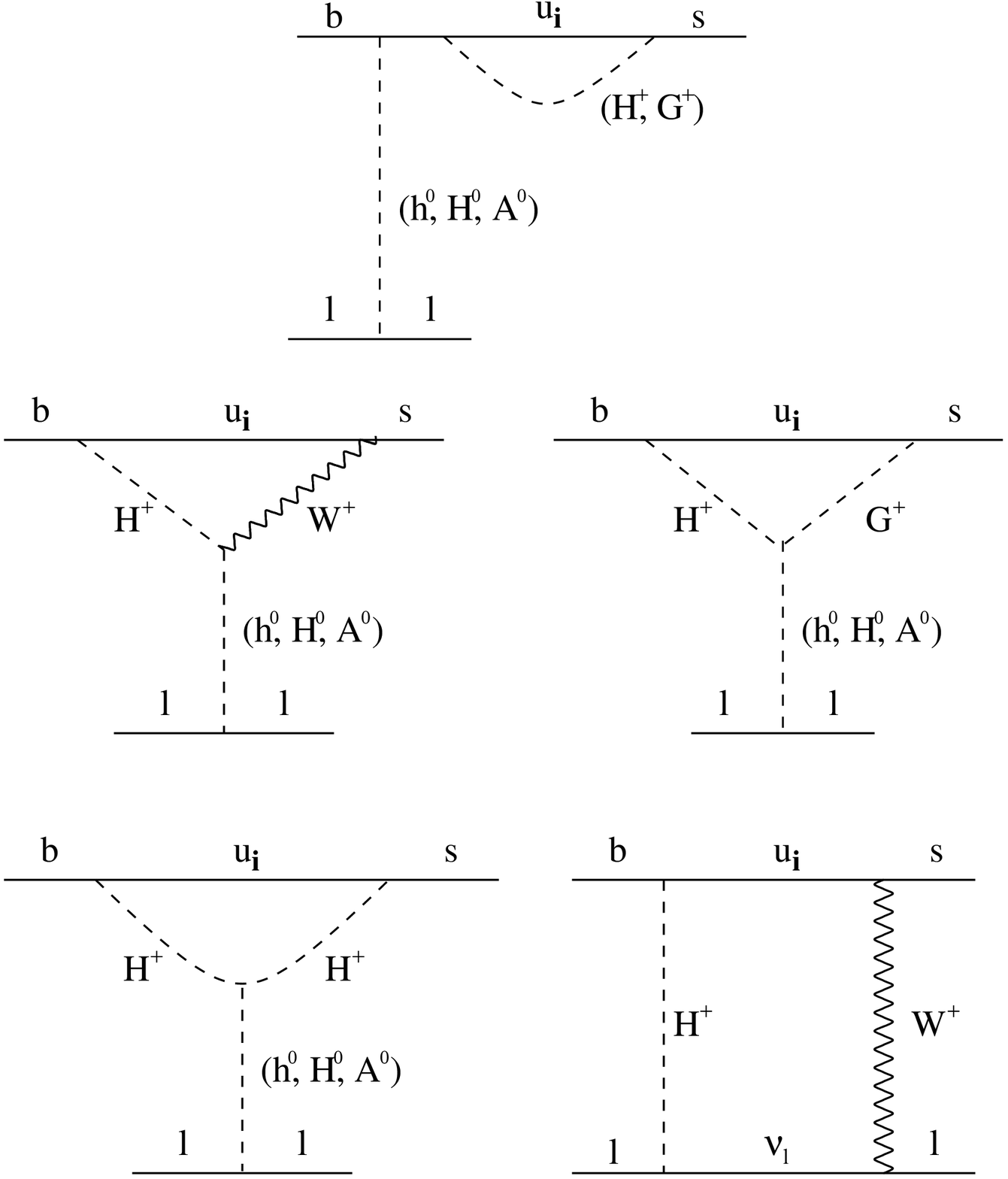,width=6.1 in}
\vskip-4cm {\huge   \caption{}}
     \end{minipage}
\end{figure}
\begin{figure}
\begin{minipage}[t]{6.1 in}
     \epsfig{file=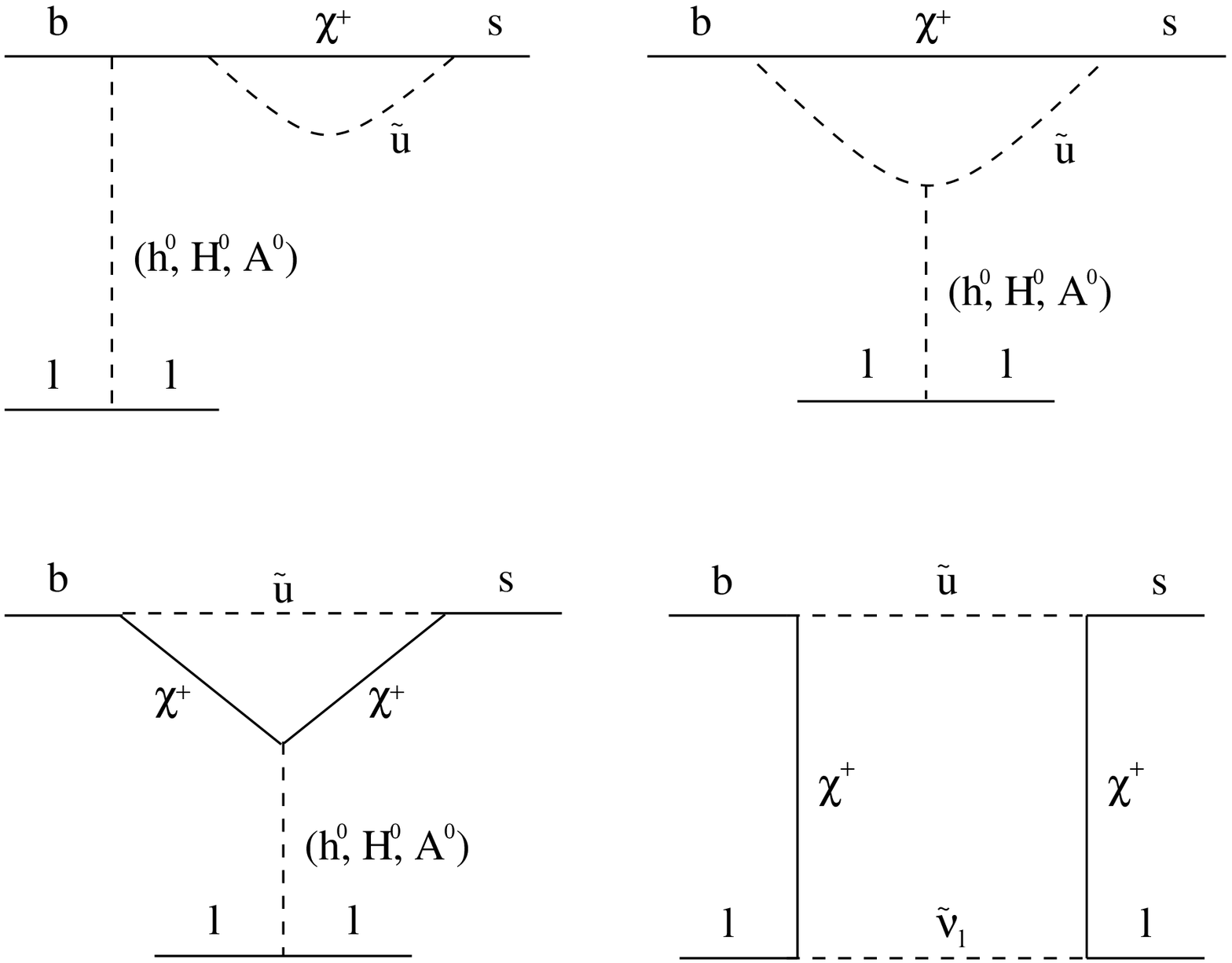,width=6.1 in}
\vskip2cm {\huge   \caption{}}
     \end{minipage}
\end{figure}
\begin{figure}
\begin{minipage}[t]{5.1 in}
\vskip 3cm     \epsfig{file=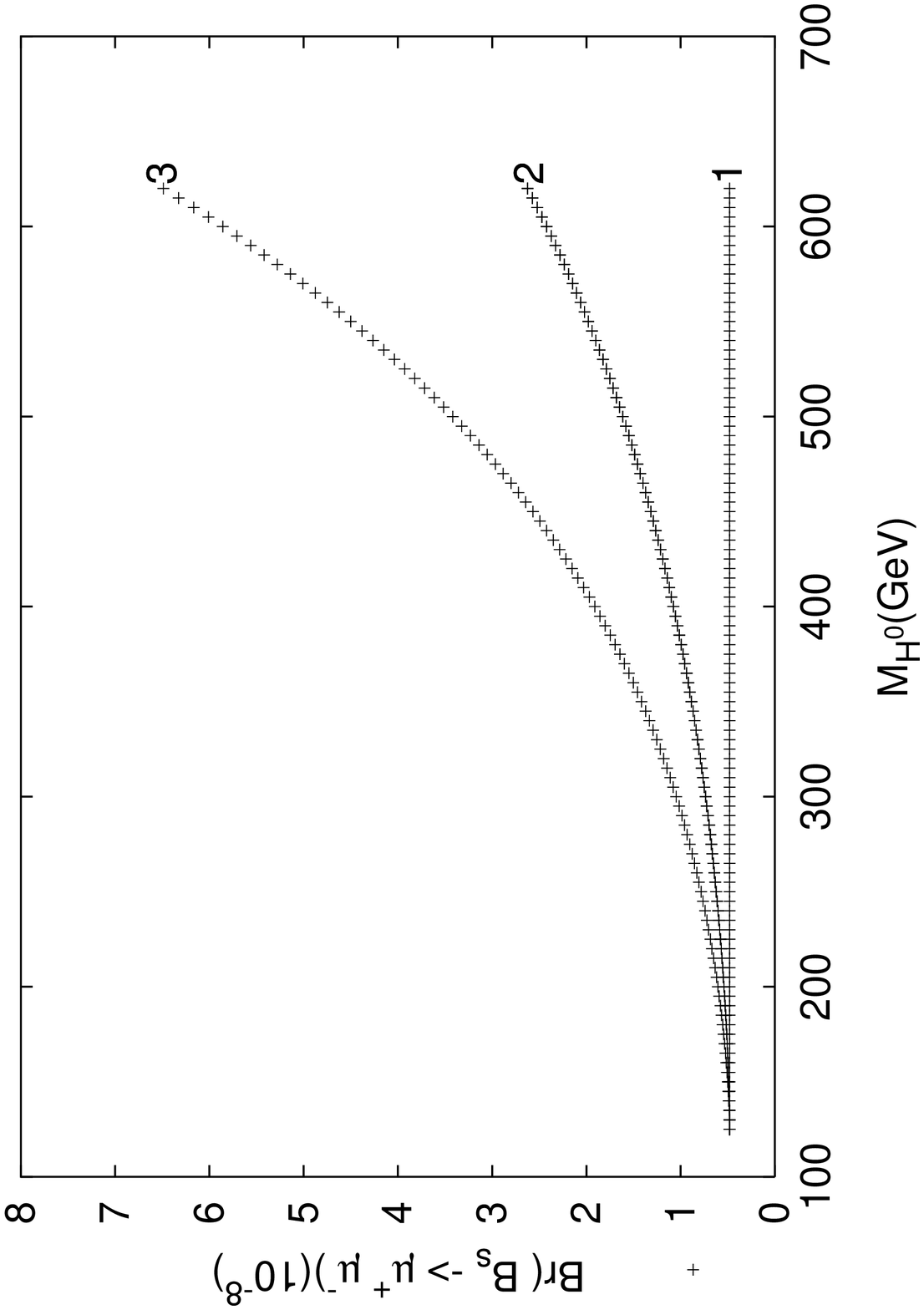}
\vskip-1cm \hspace{8cm} \caption{}
\end{minipage}
\end{figure}
\begin{figure}
\begin{minipage}[t]{5.1 in}
\vskip3cm     \epsfig{file=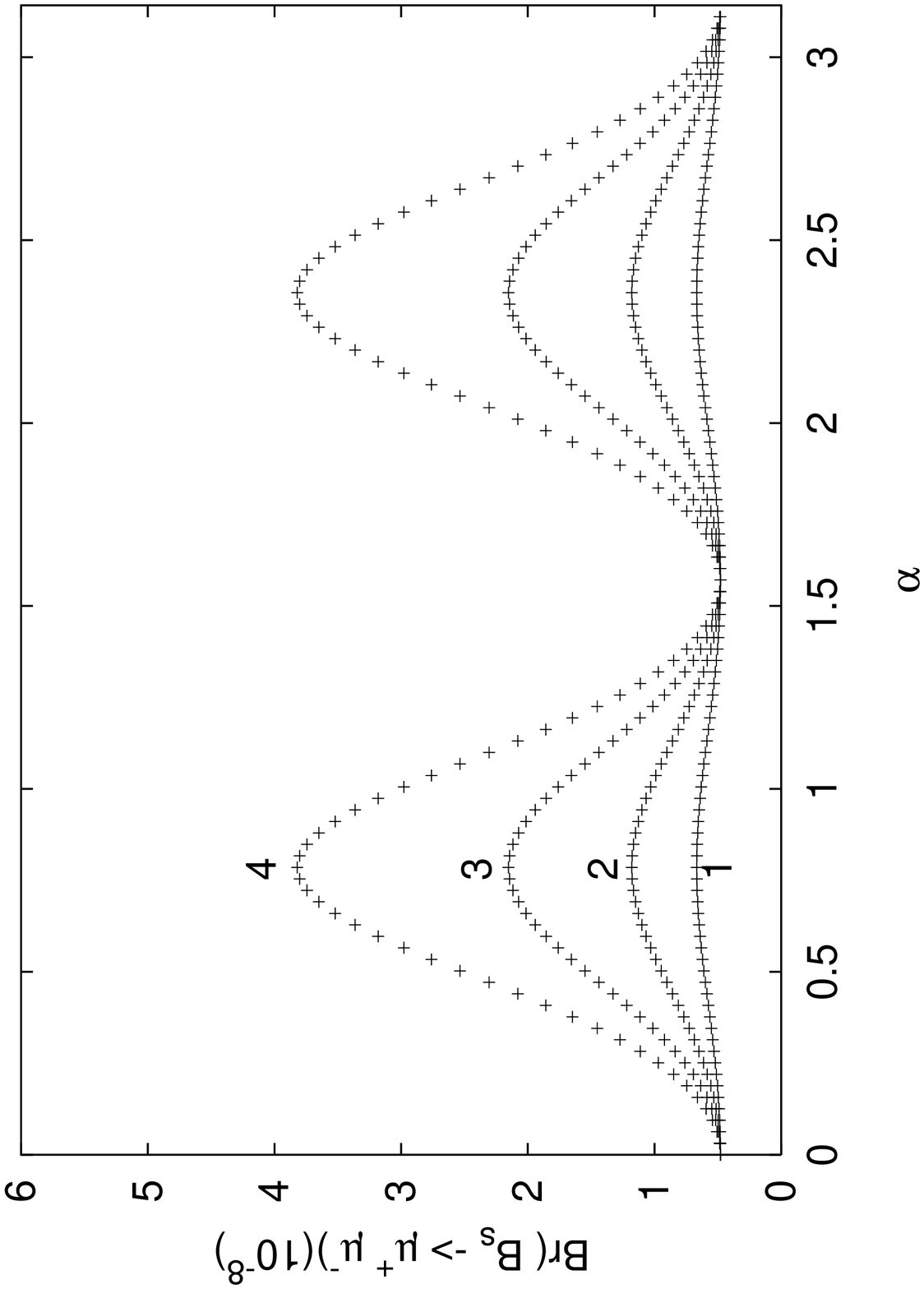}
\vskip-1cm \hspace{8cm} \caption{\huge}
\end{minipage}
\end{figure}
\begin{figure}
\begin{minipage}[t]{5.1 in}
\vskip3cm     \epsfig{file=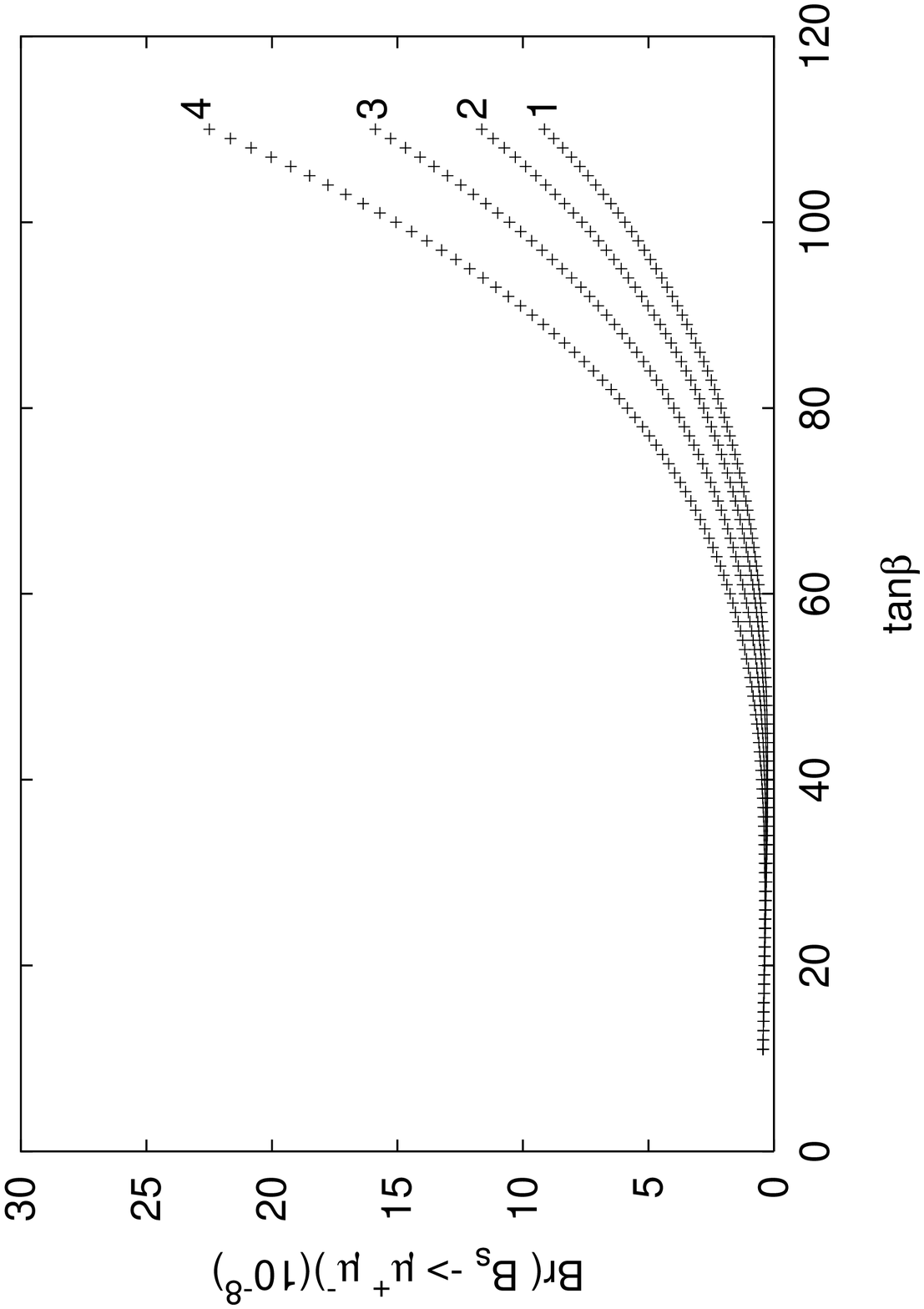}
\vskip-1cm \hspace{8cm}  \caption{}
\end{minipage}
\end{figure}
\begin{figure}
\begin{minipage}[t]{5.1 in}
\vskip-3cm     \epsfig{file=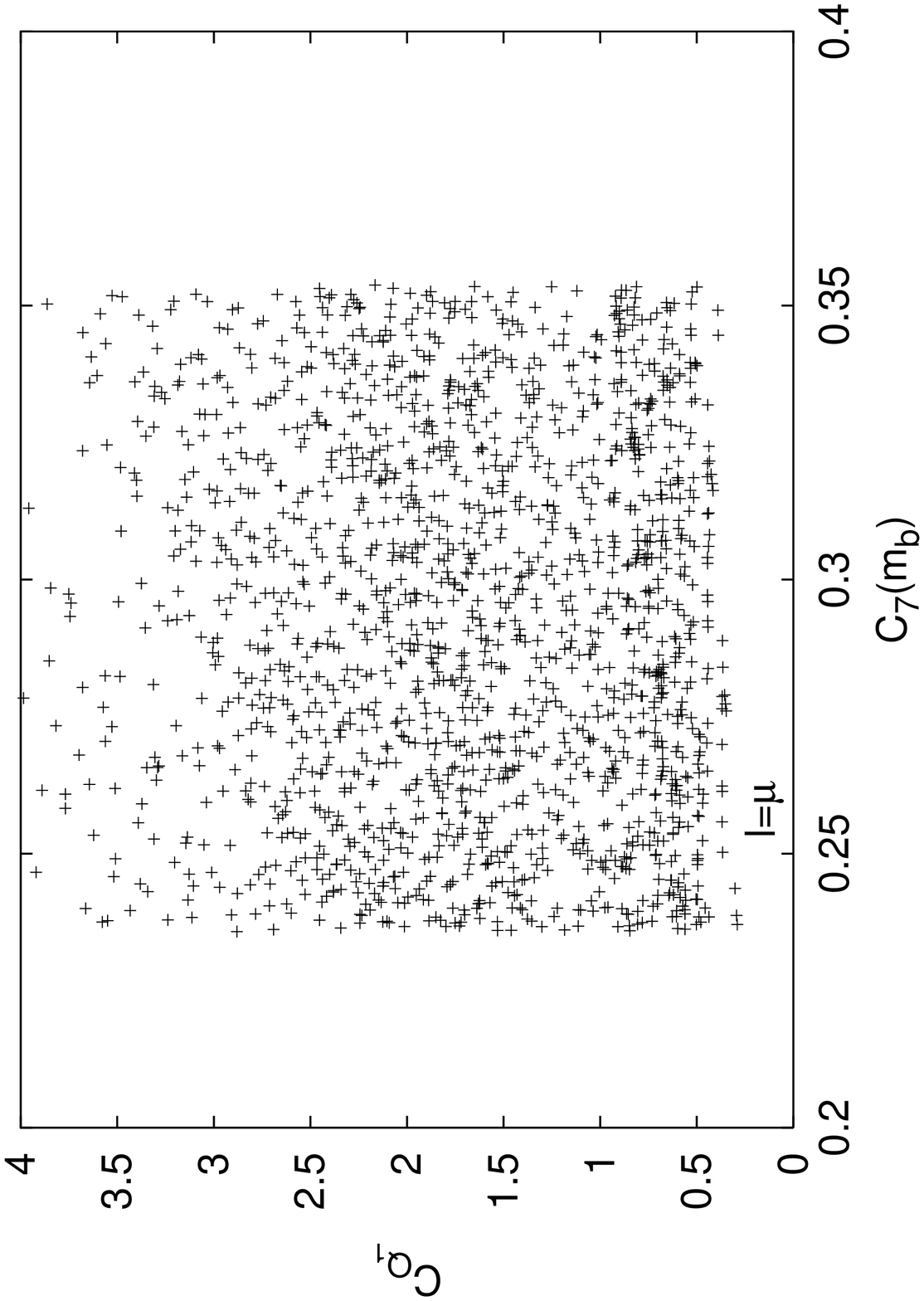}
\vskip-1cm \hspace{8cm}  \caption{}
\end{minipage}
\end{figure}
\begin{figure}
\begin{minipage}[t]{5.1 in}
\vskip-3cm     \epsfig{file=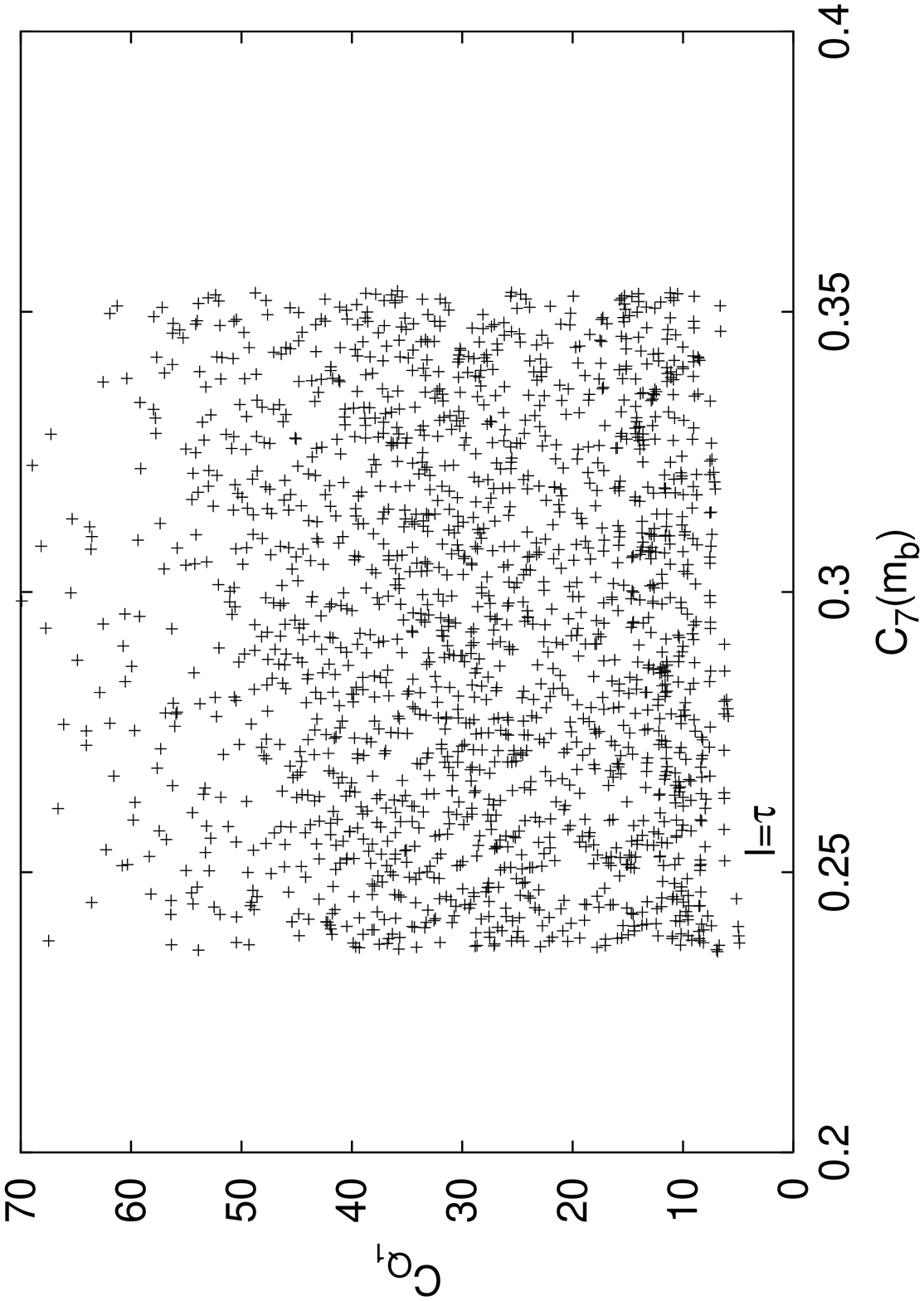}
\vskip-1cm \hspace{8cm}  \caption{}
\end{minipage}
\end{figure}
\begin{figure}
\begin{minipage}[t]{5.1 in}
\vskip-3cm     \epsfig{file=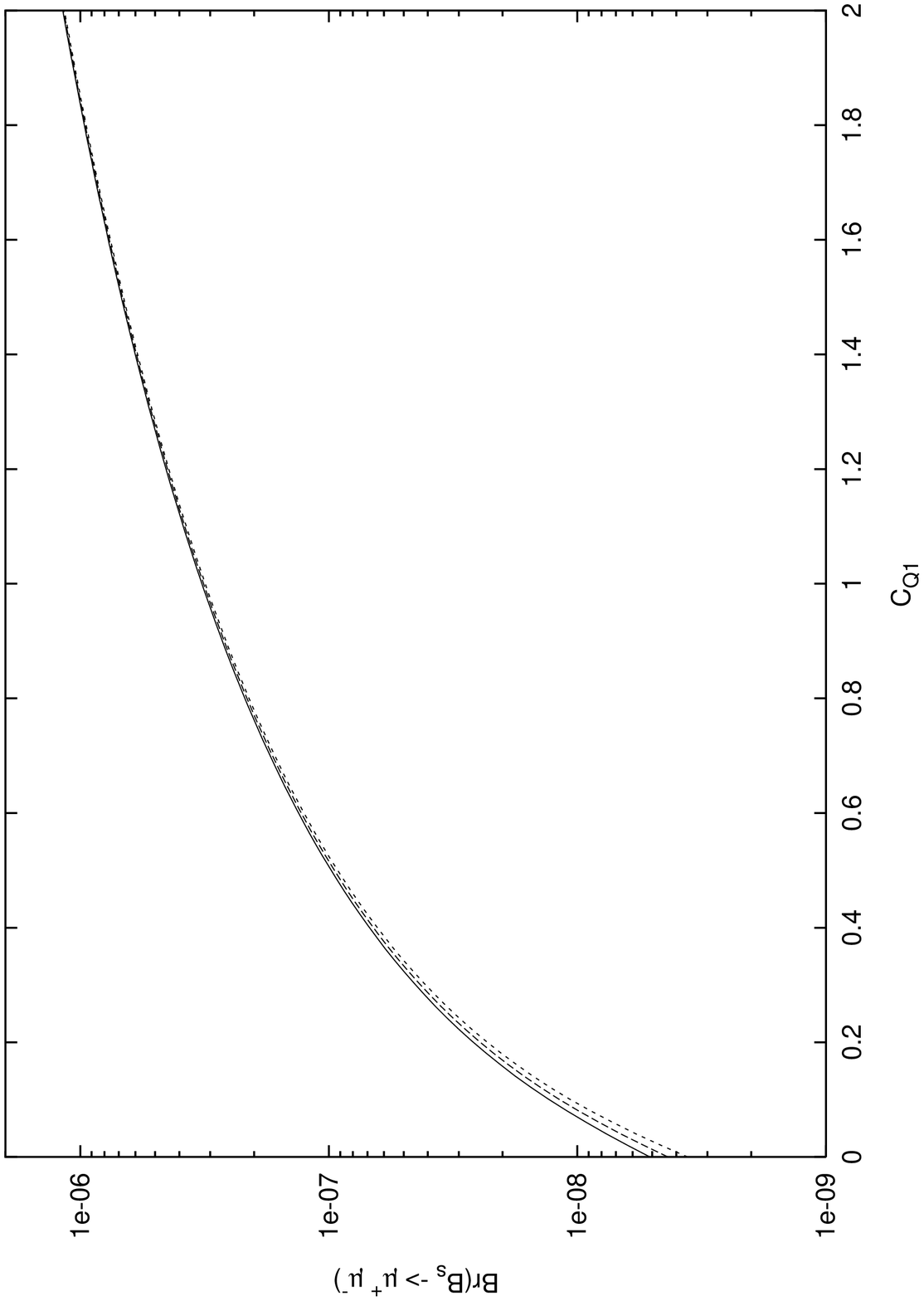}
\vskip-1cm \hspace{8cm}  \caption{}
\end{minipage}
\end{figure}
\begin{figure}
\begin{minipage}[t]{5.1 in}
\vskip-3cm     \epsfig{file=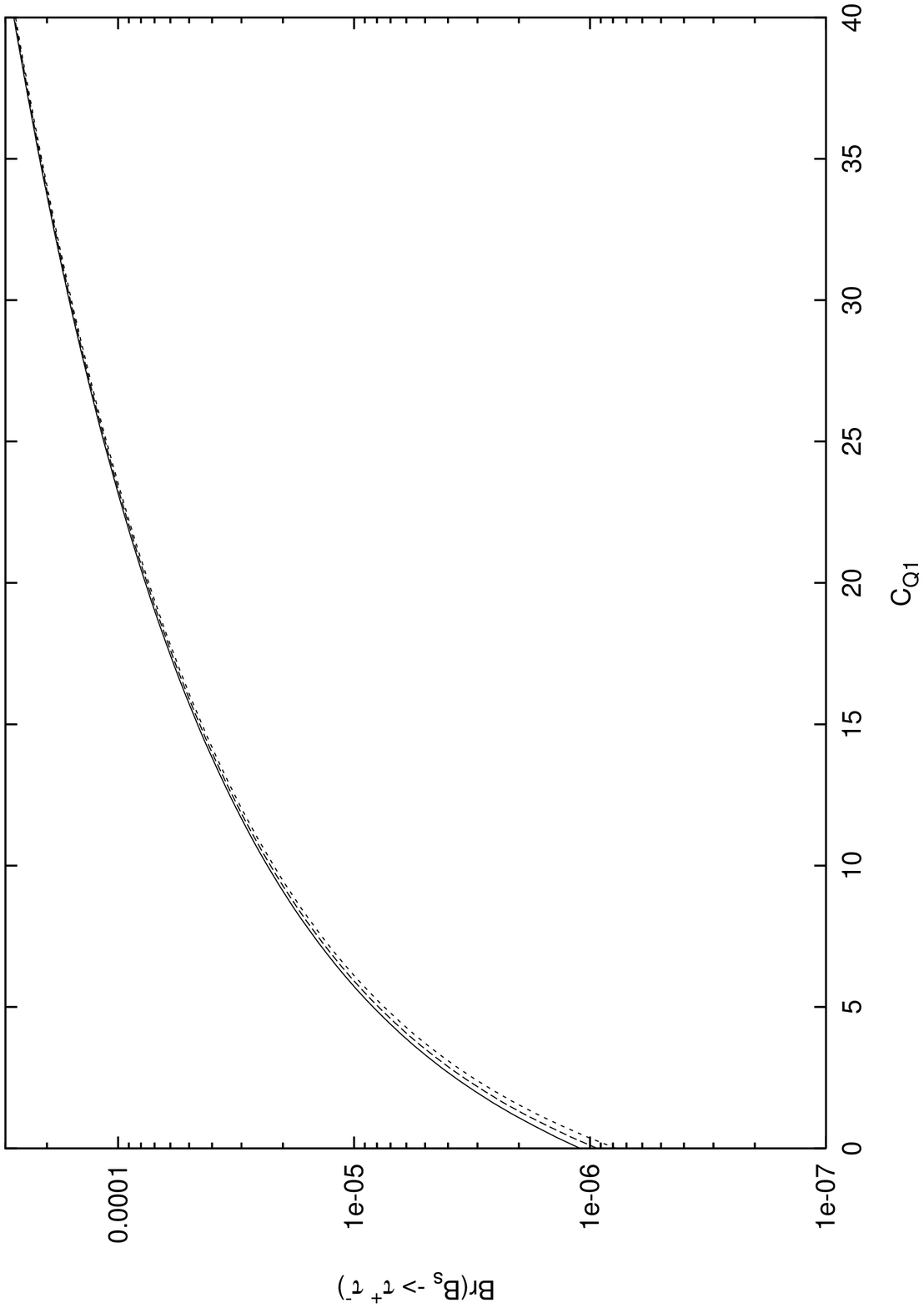}
\vskip-1cm \hspace{8cm}  \caption{}
\end{minipage}
\end{figure}


\begin{thebibliography}{99}
\bibitem{cdf} F.~Abe {\it et al.} (CDF collaboration), 
Phys.\ Rev.\ {\bf D57}, 3811 (1998).
\bibitem{ali1}
 For a recent review and complete set of refrences see A.Ali,
   DESY 7-192,hep-ph/9709507

\bibitem{hnv}
 X.~G.~He, T.~D.~Nguyen and R.~R.~Volkas,
Phys.\ Rev.\  {\bf D38} (1988) 814.

\bibitem{sk}
W.~Skiba and J.~Kalinowski, Nucl.\ Phys.\  {\bf B404} (1993) 3.

\bibitem{dhh}
Y.-B.~Dai, C.-S.~Huang and H.-W.~Huang,
Phys.\ Lett.\ {\bf B390} (1997) 257.

\bibitem{hy}
C.-S.~Huang and Q.-S.~Yan, Phys. Lett. {\bf B442} (1998) 209.
\bibitem{hly}
C.-S.~Huang, W.~Liao and Q.-S.~Yan,
Phys.\ Rev.\ {\bf D59} (1999) 011701.
\bibitem{cg}
S.~R.~Choudhury and N.~Gaur,
Phys.\ Lett.\  {\bf B451} (1999) 86.
\bibitem{ln}H.E. Logan and U. Nierste, hep-ph/0004139.
\bibitem{wise}{B.Grinstein, M.J.Savage and M.B.Wise, {\it Nucl.Phys.}{\bf B319}
(1989)271.}
\bibitem{h}Chao-Shang Huang, Commun. Theor. Phys. {\bf 2} (1983) 1265.
\bibitem{buras}G. Buchalla, A.J. Buras and M.E. Lauthenbache, Rev. Mod. Phys. {\bf 68} (1996)1125.
\bibitem{il}T.~Inami and C.~S.~Lim, Prog.\ Theor.\ Phys.\ {\bf 65}, 297 (1981) 
[erratum {\bf 65}, 1772 (1981)].
\bibitem{mas}S. Bertolini, F. Borzumati, A. Masiero and G. Ridolfi, Nucl. 
Phys. {\bf B353} (1991) 591.
\bibitem{bo}F. Borzumati, hep-ph/9310212; N. Oshimo, Nucl. Phys. {\bf
B404} (1993) 20.
\bibitem{goto}
 T.Goto,Y.Okada and Y. Shimizu {\em Phys. Rev.}{\bf D 58}(1998)094006
\bibitem{Georgi}H. Georgi, Hadronic Jour. 1 (1978) 155.
\bibitem{ghkd}
J.~F.~Gunion, H.~E.~Haber, G.~Kane and S.~Dawson,
{\it The Higgs Hunter's Guide}
(Addison--Wesley, Reading, MA, 1990); errata hep-ph/9302272.
\bibitem{bk}K.S. Babu and C. Kolda, Phys. Rev. Lett. {\bf 84} (2000) 228.
\bibitem{oyy}Y. Okada, M. Yamaguchi and T. Yanagida, Prog. Theor. Phys.
{\bf 85} (1991)1; H. Haber and R. Hempfling. Phys. Rev. Lett. {\bf 66} 1815;
J. Ellis, G. Ridolfi and F. Zwirner, Phys. Lett. {\bf B257} (1991) 83.
\bibitem{cqw}M. Carena, M. Quiros and C. Wagner, Nucl. Phys. {\bf B461} (1996) 407.
\bibitem{hcsz}C.-S. Huang and S.-H. Zhu, Phys. Rev. {\bf D60} (1999) 075012.
\bibitem{bsg4}R. Garisto and J.N. Ng Phys. Lett. B 315 (1993) 372.
\bibitem{fe}K. Pitts, Proceedings 3th Workshop on Heavy Quarks at Fixed Target(HQ98), Batavia, USA.
\bibitem{ahmed} S. Ahmed et.al, CLEO collaboration, CLEO CONF99-10, hep-ex/9908022.
\end{thebibliography}
\end{document}